\documentclass[a4paper,11pt]{article}
\pdfoutput=1 

\usepackage{jcappub} 

\usepackage[T1]{fontenc} 
\usepackage{txfonts}
\usepackage{braket}
\usepackage{latexsym}
\usepackage{wrapfig}
\usepackage{tabularx}
\usepackage{comment}
\newcommand{\irphi}{{\phi^<}}

\newcommand{\phidelta}{\phi_\Delta}

\newcommand{\irphid}{{\phi_\Delta^<}}
\newcommand{\irphic}{{\phi_c^<} }
\newcommand{\ddirphic}{{\ddot \phi_c^<} }
\newcommand{\dirphic}{{\dot \phi_c^<} }
\newcommand{\irvc}{{v_c^<}}
\newcommand{\dirvc}{{\dot v_c^<}}
\newcommand{\irvd}{{v_\Delta^<}}
\newcommand{\uvvc}{{v_c^>}}
\newcommand{\uvvd}{{v_\Delta^>}}
\newcommand{\uvphid}{{\phi_\Delta^>}}
\newcommand{\uvphic}{{\phi_c^>}} 
\newcommand{\volint}[1]{\int\mathrm{d}^4\!{#1}}
\newcommand{\modeint}[1]{\int\frac{\mathrm{d}^3 k_{#1}}{(2\pi)^3}}

\newcommand{\pathint}{\int \mathcal{D}}

\newcommand{\joah}[3]{j_0\left(\ah{#1}\left|\vec{#2}-\vec{#3}\right|\right)}
\newcommand{\der}{\partial}
\newcommand{\ah}[1]{\epsilon a(t_{#1})H}
\newcommand{\no}{\nonumber}
\newcommand{\modefn}[3]{\Phi_{k_{#1}}(#2)\Phi^*_{k_{#1}}(#3)}

\newcommand{\immodefn}[1]{\epsilon e^{H\Delta {#1}}\cos(\epsilon e^{H\Delta {#1}})-\sin(\epsilon e^{H\Delta {#1}})}
\newcommand{\remodefn}[1]{\epsilon e^{H\Delta {#1}}\sin(\epsilon e^{H\Delta {#1}})+\cos(\epsilon e^{H\Delta {#1}})}
\def\para{%
\setlength{\unitlength}{1pt}%
\thinlines 
\begin{picture}(12, 12)
\put(0,0){/}
\put(2,0){/}
\end{picture}
}

\title{Statistical nature of infrared dynamics on de Sitter background}

\author[a]{Junsei Tokuda}
\author[a,b]{and Takahiro Tanaka}

\affiliation[a]{Department of Physics, Kyoto University,
\\Kyoto 606-8502, Japan}
\affiliation[b]{Center for Gravitational Physics, Yukawa Institute for Theoretical Physics, Kyoto University,\\Kyoto 606-8502, Japan}

\emailAdd{tokuda@tap.scphys.kyoto-u.ac.jp}
\emailAdd{t.tanaka@tap.scphys.kyoto-u.ac.jp}

\abstract{
In this study, we formulate a systematic way of deriving an effective equation of motion(EoM) for long wavelength modes of a massless scalar field with a general potential $V(\phi)$ on de Sitter background, and investigate whether or not the effective EoM can be described as a classical stochastic process. Our formulation gives an extension of the usual stochastic formalism to including sub-leading secular growth coming from the nonlinearity of short wavelength modes. Applying our formalism to $\lambda \phi^4$ theory, we explicitly derive an effective EoM which correctly recovers the next-to-leading secularly growing part at a late time, and show that this effective EoM can be seen as a classical stochastic process. Our extended stochastic formalism can describe all secularly growing terms which appear in all correlation functions with a specific operator ordering. The restriction of the operator ordering will not be a big drawback because the commutator of a light scalar field becomes negligible at large scales owing to the squeezing.}

\begin{document}
{\baselineskip0pt
\rightline{\baselineskip16pt\rm\vbox to-20pt{
           \hbox{YITP-17-84, KUNS-2697}
\vss}}%
}

\maketitle
\flushbottom

\section{Introduction}
Inflationary universe scenario is one of the leading paradigm in modern cosmology \cite{Staro1980,Guth1981,Linde1982,Linde1983}, and has been studied by many researchers from various aspects. One of the most important outcome of the inflationary paradigm is the generation of primordial cosmological perturbations, originating from quantum fluctuations of fields. Our main interest is in long-wavelength modes of quantum fields well outside the horizon scale, which we simply call IR modes here, at the end of inflation, as they generate the observed perturbations, {\it e.g.}, in cosmic microwave background (CMB). 
The theoretical framework of the calculation of correlation functions of primordial perturbations during inflation is not fully justified yet because of the 
issues about ``IR divergences'' \cite{Urakawa2009-1, Urakawa2009-2, Seery2010, Tanaka2013}. It is known that correlation functions are IR divergent in the theory of a minimally coupled massless scalar field on de Sitter background, which mimics isocurvature perturbations during inflation. 
The existence of such light degrees of freedom other than inflaton seems to be required by the fundamental physics, such as the string theory \cite{Baumann2015}. Even if we introduce an IR cutoff in comoving momentum space, loop corrections generate IR secular growth which invalidates the perturbative QFT computations at a late time, and these loop corrections depend on the artificial IR cutoff. Therefore, simple QFT expectation values cannot be considered as observables. Thus, we need a consistent prescription of calculating observables, which is free from IR divergences. 

It is known that leading-order(LO) IR secular growth terms, 
which mean the most rapidly secularly growing pieces at each order of the expansion with respect to the coupling constant, 
are described as Brownian motion \cite{Tsamis2005, Finelli2009, Finelli2010, Onemli2015}. 
At this level of approximation, one can describe the inflationary dynamics 
as if it were a classical stochastic process \cite{Starobinsky1982, Staro1986, Nakao1988, Nambu1989, Staro1994}, which is called the stochastic approach. 
If we interpret that the stochastic approach allows us 
to assign the probabilities to various classical realizations of fields, 
there arises the picture of the eternal inflation to describe the very early universe \cite{Linde1986}. 
This interpretation also applies to calculate correlation functions of adiabatic perturbations \cite{Fujita2013, Fujita2014, Vennin2015, Assa2016} and gives us a 
prescription of calculating observables, which are free from LO IR 
secular growth.\footnote{In evaluating the correlation functions of the fluctuations of our observable universe based on this picture, one starts the stochastic dynamics from one Hubble patch at the time when the mode corresponding to the largest observable scale exceeds the Hubble scale. This allows us to neglect deep-IR modes beyond the observable scales in calculating observable correlation functions.} 
Once we accept this interpretation, the next question is whether or not one 
can compute \emph{all} IR correlation functions 
in a good approximation by extending this stochastic approach.\footnote{Recently several papers appeared which try to extend Stochastic Formalism, in Schr$\rm{\ddot{o}}$dinger picture \cite{Collins2017}, or by using path integral method \cite{Moss2016}.}
If this extension fails seriously, the current picture of inflationary universe might be drastically modified. 

Motivated by the above observations, we investigate how to derive an 
effective EoM for IR modes which correctly describes all IR correlation functions. As a first step, we focus on the case of a minimally coupled massless scalar field with a general potential 
$V(\phi)$ on de Sitter background in this study. This paper is organized as follows. In sec.~\ref{IRlog}, we show the IR secular growth explicitly in $\lambda\phi^4$ theory on de Sitter background as an example. In sec.~\ref{Genstoc}, we establish how to derive an effective EoM for IR modes which includes all IR secular growth in the theory of a minimally coupled massless scalar field with a general potential $V(\phi)$. 
In sec.~\ref{systemNLO}, we propose a systematic way of estimating the order of the IR secular growth for each diagram which constitutes the effective action for IR modes, and derive an effective EoM which is valid up to next-to-leading order (NLO) IR secular growth by applying the proposed method to $\lambda\phi^4$ theory. We show that this NLO stochastic EoM can be interpreted as a classical stochastic process. 
Finally in sec.~\ref{concl}, we summarize our results. Several details of calculations are presented in Appendices.
We adopt natural units,
\begin{equation*}
c=1=\hbar\,,\quad M_\mathrm{pl}\coloneqq\sqrt{\frac{1}{8\pi G}}=1\,.
\end{equation*}

\section{IR secular growth and stochastic formalism}\label{IRlog}

In this section, we briefly recapitulate the so-called "stochastic formalism"
that describes LO IR secular growth correctly 
\cite{Staro1986, Staro1994, Tsamis2005}\footnote{This stochastic formalism is also called Starobinsky's stochastic formalism.}.

\subsection{IR secular growth}\label{IRsec}
As a concrete model, we consider a weakly self-interacting scalar field 
on de Sitter background in flat chart $(t,\vec x)$ (or $(\eta,\vec x)$ ). 
More precisely, the action and the background spacetime are given by 
\begin{subequations}
\begin{align}
S&=\volint{x}\,a^3\left(-\frac{1}{2}(\der\phi)^2-\frac{\lambda}{4!}\phi^4\right),\quad 0<\lambda\ll1\,,\\
ds^2&=-dt^2+a^2(t)\delta_{ij}dx^idx^j=a^2(\eta)\left(-d\eta^2+\delta_{ij}dx^idx^j\right)\,,\\
a(t)&=e^{Ht}=-\frac{1}{H\eta}\,,\qquad H\coloneqq\frac{\dot a}{a}=\mathrm{constant}\,.
\end{align}
\label{eq:setup}
\end{subequations}
Here, dot denotes the derivative with respect to $t$. For brevity, we abbreviate the argument $t$, if it does not cause any confusion. 
We introduce IR modes 
$\irphi$ defined by
\begin{align}
\irphi(x) \coloneqq \modeint{}\Theta\left(\epsilon a H-k\right)\phi_{\vec k}(t)e^{i\vec k\cdot\vec x}\,,\qquad k\coloneqq\bigl|\,\vec k \, \bigr|\,,\label{cg1}
\end{align}
where $\Theta(z)$ is the Heaviside function and $\epsilon$ is a non-dimensional small parameter. 
Roughly speaking, $\irphi(\vec x,t)$ is a coarse-grained field 
averaged over the comoving length scale $1/\epsilon aH$ on a constant time slice. 
We introduce an IR comoving momentum cutoff
\begin{align}
k_0&\coloneqq\epsilon a_0H\,,
\end{align}
where $a_0$ is 
the value of $a$ at the initial time $t_0$ 
and neglect all modes with a momentum below this cutoff, to 
avoid the IR divergences due to the momentum integral. 
Our expectation is 
that the initial time can be smoothly sent to the past infinity, 
once an appropriate effective EoM for IR modes is derived. 
Under this setup correlation functions of $\irphi(\vec x,t)$ have secularly growing terms, {\it e.g.}  
\begin{align}
\bra0\irphi^{2n}(x)\ket0
&=\sum_{l=0}^\infty \sum_{m=0}^{2l+n-1}c_{lm}
   \lambda^l\left(\ln\frac{a}{a_0}\right)^{2l+n-m}
+ \mbox{(IR regular part)}
\nonumber\\
&\sim H^2\left(\ln\frac{a}{a_0}\right)^n+\lambda H^2\left(\left(\ln\frac{a}{a_0}\right)^{n+2}+\left(\ln\frac{a}{a_0}\right)^{n+1}+\cdots\right)+\mathcal{O}\left(\lambda^2\right)\,,\label{corfn1}
\end{align}
where $\ket0$ is the state vector corresponding to the non-interacting Bunch-Davies vacuum at the initial time. 
We refer to $m=0$ terms in eq.~\eqref{corfn1}, 
which represent the LO IR secular growth, as LO IR terms. 
Similarly, we refer to $m=1$ terms as NLO IR terms. 
In general, $m=m_1$ terms are denoted by $\mathrm{N^{m_1} LO}$ IR terms.  
Since the terms higher order in $\lambda$ 
get larger after passing the critical time $t_\mathrm{np}$ 
determined by the condition 
$
\lambda(\ln{a(t_\mathrm{np})}/{a_0})^2\approx 1\,,
$ 
which implies
\begin{equation*}
 t_\mathrm{np}-t_0\approx \frac{1}{\sqrt{\lambda}H}\,,
\end{equation*}
perturbative expansion with respect to $\lambda$ breaks down after $t_\mathrm{np}$. 

\subsection{Stochastic formalism}\label{stoc}
The stochastic EoM for a canonical scalar field 
with a general potential $V(\phi)$ is given as~\cite{Staro1986, Staro1994, Tsamis2005}
\begin{align}
\label{LOstoc}
\begin{split}
{\dot\phi}^<(x)&=-\frac{1}{3H}V'\left(\irphi(x)\right)+\xi_\phi(x)\,,
\\
\left<\xi_\phi(x_1)\xi_\phi(x_2)\right>
&=\frac{H^3}{4\pi^2}\delta(t_1-t_2)\,
 \Theta\left(\frac{1}{\ah{1}}-\left|\vec{x}_1-\vec{x}_2\right|\right)\,,
\end{split}
\end{align}
where $\left<\cdots\right>$ means the statistical ensemble average. 
This equation is obviously equivalent to an EoM of Brownian 
motion under an external force $V'(\phi)$, 
and means that each horizon patch evolves independently. 
Therefore, this equation can be solved non-perturbatively 
with respect to $\lambda$, which 
makes the obtained solution valid even at a late epoch where 
perturbative QFT calculations fail to converge.
It has been checked explicitly that  
the stochastic EoM~\eqref{LOstoc} correctly recovers LO IR terms 
(See {\it e.g.} \cite{Tsamis2005}). 
Since eq.~\eqref{LOstoc} is identical to an equation that 
describes a stochastic process in classical mechanics,
it might be allowed to interpret that the stochastic EoM is 
the one that attributes probabilities to various possible realizations of the
classical time evolution. 
In this stochastic interpretation, the LO IR secular growth is simply
interpreted as an increase of the statistical variance. 

However, one should recall that the stochastic EoM~\eqref{LOstoc}
neglects sub-LO IR terms, which also increase indefinitely in time.  
Furthermore, what we are interested in would be the finite part of the 
correlation functions that after the secular growth effects are removed. 
Therefore, there seems to be no reason why one can be satisfied 
with the treatment neglecting the 
sub-LO IR terms. It seems non-trivial whether or not an effective EoM 
which can describe all IR correlation functions with sufficient precision can be obtained in the form similar 
to eq.~\eqref{LOstoc}. 

\section{Extended stochastic formalism}\label{Genstoc}
In this section, we give a prescription 
to derive an effective EoM for IR modes which correctly recovers 
all IR correlation functions. 

\subsection{Assumptions}\label{setup}
We consider the system defined by the  following Hamiltonian density
\begin{subequations}
\label{hami}
\begin{align}
\mathcal{H}[v,\phi]&=\mathcal{H}_0[v,\phi]+V(\phi)\,, \\
\mathcal{H}_0[v,\phi]&\coloneqq\frac{1}{2}v^2+\frac{(\nabla\phi)^2}{2a^2}\,,\label{hami1}
\end{align}
\end{subequations}
where we introduced $v$ as
\begin{equation*}
v\coloneqq\frac{1}{a^3}\Pi_\phi=\dot\phi\,,
\end{equation*}
and $\Pi_\phi$ is the ordinary conjugate momentum of $\phi$. The interaction potential $V(\phi)$ includes the mass term $m^2\phi^2$, and hence the non-interacting Hamiltonian $\mathcal{H}_0$ is the one for an exactly massless scalar field.

We make two assumptions on the initial state.
\begin{enumerate}
\item Assuming 
that the potential $V(\phi)$ is turned on at $t=t_0$, 
we take the Bunch-Davies vacuum state for a free field at $t=t_0$. 
The Bunch-Davies vacuum state $\ket0$ is specified by the conditions
\begin{equation}
\hat a_{\vec k}\ket 0=0\,,
\end{equation}
when the interaction picture fields $\phi_\mathrm{I}$, 
$v_\mathrm{I}$ are expanded as
\begin{align}
\hat\phi_\mathrm{I}(x)&=\modeint{}\left[\Phi_k(t)e^{i\vec k\cdot\vec x}\hat a_{\vec k}+(\mathrm{h.c.})\right]\,, \label{quantize1}\qquad
\hat v_\mathrm{I}(x)=\modeint{}\left[\dot\Phi_k(t)e^{i\vec k\cdot\vec x}\hat a_{\vec k}+(\mathrm{h.c.})\right]\,,
\end{align}
where $(\mathrm{h.c.})$ stands for hermitian conjugate and 
\begin{align}
\Phi_k(t)&=\frac{H}{\sqrt{2k^3}}\left(1+ik\eta \right)e^{-ik\eta}\,.
\label{quantize2}
\end{align}
These mode functions are properly normalized to satisfy 
\begin{align}
 \dot\Phi_k(t) \Phi_k^*(t)-\Phi_k (t) \dot\Phi_k^*(t)=\frac{-i}{a^3}\,, 
\label{eq:normalization}
\end{align}
and the creation and annihilation operators 
$\hat a_{\vec k}$ and ${\hat{a}}^\dagger_{\vec{k}}$ satisfy 
the commutation relations
\begin{align}
\left[\,{\hat a}_{\vec k}\,,\,\hat{a}_{\vec k'}{}^{\!\!\!\!\!\dag}\,\right]
&=(2\pi)^3\delta^{(3)}(\vec{k}-\vec{k}')\,,\qquad
\left[\,\hat a_{\vec k}\,,\, {\hat{a}}_{\vec{k}'}\,\right]=0\,,\qquad
\left[\,\hat a_{\vec k}{}^{\!\!\!\dag}\,, \,\hat{a}_{\vec k'}{}^{\!\!\!\!\!\dag}\,\right]=0\,. 
\end{align}

\item We neglect modes that are already belonging to IR modes 
$\phi^<(x)$ and $v^<(x)$ 
at the initial time $t=t_0$, where $v^<(x)$ is defined 
in the same manner as $\phi^<(x)$ in eq.~\eqref{cg1}. 
This is equivalent to introducing an IR cutoff 
$k_0\coloneqq\epsilon a_0H$ to the comoving momentum $k$. 
Namely, the range of integration in eq.~\eqref{quantize1} is 
restricted to $k>k_0$. 
Since the IR divergences associated with the momentum integral are 
removed by this prescription, 
one can discuss correlation functions 
within the standard framework of QFT.
\end{enumerate}

Since all modes are belonging to UV modes at the initial time, 
each IR mode has the crossing time transferred from a UV mode. 
This crossing time is given by 
\begin{equation*}
t_c(k)\coloneqq\frac{1}{H}\ln\frac{k}{\epsilon H}\,.
\end{equation*}
As usual, the interaction picture fields are introduced as 
\begin{align}
\label{hisenint}
\phi_\mathrm{H}(\vec x,t)&=U^\dagger(t,t_0)\phi_\mathrm{I}(\vec x,t)U(t,t_0)\,, \quad v_\mathrm{H}(\vec x,t)=U^\dagger(t,t_0)v_\mathrm{I}(\vec x,t)U(t,t_0)\,, 
\end{align}
with 
\begin{align}
U(t,t_0)&\coloneqq e^{-i\int^t_{t_0}\mathrm{d}t\,H_{\mathrm{int}}}\,.
\end{align}
We decompose the Heisenberg picture fields into two parts, {\it i.e.}
UV modes and IR modes as
\begin{align}
\phi_\mathrm{H}&=\phi^>_\mathrm{H}+\phi^<_\mathrm{H}\,,
\end{align}
where
\begin{align}
\phi_\mathrm{H}^>(\vec{x},t) &\coloneqq \modeint{} \Theta(k-\epsilon aH)\ 
\phi_k(t)e^{i\vec k\cdot \vec x}\,.
\end{align}
Similarly, $v_\mathrm{H}$ is also decomposed. 
This decomposition is based only on the comoving momentum. 
Therefore, if we decompose the interaction picture fields as 
$\phi_\mathrm{I}=\phi_\mathrm{I}^>+\phi_\mathrm{I}^<$ 
and $v_\mathrm{I}=v_\mathrm{I}^>+v_\mathrm{I}^<$ in the same manner, 
the relations \eqref{hisenint} hold for UV and IR modes, respectively. 
\subsection{Schwinger-Keldysh formalism and 
splitting between IR and UV modes}\label{pert}

In order to derive an effective EoM for IR modes, 
we integrate out UV modes.\footnote{The idea itself that deriving an effective EoM for IR modes by integrating out UV modes is already proposed in \cite{Morikawa1990}.}
We start with the path integral expression in 
the Schwinger-Keldysh formalism 
(for a review, see \cite{Kamenev2009}). 
The generating functional is written as 
\begin{align}
Z[J_{\pm}]
&=\pathint\phi_{\pm}\mathcal{D}v_{\pm}\mathrm{exp}\left[i\volint{x} \, a^3
\left(v_+\dot\phi_+-\frac{1}{2}v_+^2-\frac{1}{2}\frac{(\nabla\phi_+)^2}{a^2}-V(\phi_+)+J^{\phi}_+\phi_++J^{v}_+v_+\right)-(+\leftrightarrow-)\right]\label{generate}\,,
\end{align}
with boundary conditions
\begin{equation}
\phi_+(x)=\phi_-(x)\,, \quad v_+(x)=v_-(x) \quad \mathrm{for} \quad t=t_\mathrm{max}\,,
\end{equation}
where $t_\mathrm{max}$ is an appropriately chosen 
maximum time. 
In the Keldysh basis $(v_c,\phi_c,v_\Delta,\phi_\Delta)$, which is defined by
\begin{align}
\phi_c&\coloneqq\frac{\phi_++\phi_-}{2}\,, \quad  v_c\coloneqq\frac{v_++v_-}{2}\,,\\
\phi_\Delta&\coloneqq\phi_+-\phi_- \,,\quad v_\Delta\coloneqq v_+-v_-\,,
\end{align}
the generating functional is rewritten as
\begin{align}
\label{hamiact}
\begin{split}
Z[J_c,J_\Delta]&=\pathint\phi_{c,\Delta}\mathcal{D}v_{c,\Delta}
\, e^{iS_\mathrm{H}[v_c,v_\Delta,\phi_c,\phi_\Delta]+\volint{x}\,a^3 \left(J_c^\phi(x)\phi_\Delta(x)+J_\Delta^\phi(x)\phi_c(x)+J_c^v(x)v_\Delta(x)+J_\Delta^v(x)v_c(x)\right)}\,,\\
S_\mathrm{H}[v_c,v_\Delta,\phi_c,\phi_\Delta]&\coloneqq S_\mathrm{H,0}[v_c,v_\Delta,\phi_c,\phi_\Delta]+S_\mathrm{H,int}[\phi_c,\phi_\Delta]\,,\\
S_\mathrm{H,0}[v_c,v_\Delta,\phi_c,\phi_\Delta]&
 \coloneqq\volint{x}\,a^3
 \left(v_c\dot\phi_\Delta+v_\Delta\dot\phi_c-v_c v_\Delta
   -\frac{(\vec\nabla\phi_c)\,(\vec\nabla\phi_\Delta)}
         {a^2}\right)\,,\\
S_\mathrm{H,int}[\phi_c,\phi_\Delta]&\coloneqq-\volint{x}\,a^3
\left(V\left(\phi_c+\frac{1}{2}\phi_\Delta\right)
  -V\left(\phi_c-\frac{1}{2}\phi_\Delta\right)\right)\,,
\end{split}
\end{align}
with boundary conditions
\begin{equation}
\phidelta(x)=0 \,,\quad  v_\Delta(x)=0 \quad \mathrm{for} \quad t=t_\mathrm{max}\,.
\end{equation}

Next, we split the path integral 
into two parts corresponding to UV modes and IR modes. 
Focusing on the non-interacting part, we neglect the interaction term 
$V(\phi)$ for a while. 
Then, the above path integral is decomposed as 
\begin{align}
\pathint\phi_{c,\Delta}\mathcal{D}v_{c,\Delta}~e^{iS_\mathrm{H,0}}
=\pathint\phi^<_{c,\Delta}\,\mathcal{D}v^<_{c,\Delta}~  
e^{iS^<_\mathrm{H,0}}\pathint\phi^>_{c,\Delta}
\,\mathcal{D}v^>_{c,\Delta}~e^{iS^>_\mathrm{H,0}}\, e^{i\tilde S_\mathrm{bilinear}}\,,\label{funcdiv}
\end{align}
where
\begin{align*}
S^<_\mathrm{H,0}&=S^<_\mathrm{H,0}\left[\irvc,\irvd,\irphic,\irphid\right]\coloneqq S_\mathrm{H,0}\left[v_c=\irvc,v_\Delta=\irvd,\phi_c=\irphic,\phi_\Delta=\irphid\right]\,,\\
S^>_\mathrm{H,0}&=S^>_\mathrm{H,0}\left[v^>_c,v^>_\Delta,\uvphic,\uvphid\right]\coloneqq S_\mathrm{H,0}\left[v_c=\uvvc,v_\Delta=\uvvd,\phi_c=\uvphic,\phi_\Delta=\uvphid\right]\,,\\
\tilde S_\mathrm{bilinear}&=\tilde S_\mathrm{bilinear}\left[\irvd,\irphid,\uvvc,\uvphic\right]\,.
\end{align*}
The term $e^{i\tilde S_\mathrm{bilinear}}$ 
represents the transition from UV modes to IR modes 
due to the cosmic expansion. 
To understand the necessity of 
this bilinear interaction term, 
let us consider a propagator with a given comoving momentum $\vec k$ with endpoints at
$t_1$ and $t_2$, and suppose $k$ satisfies $
\ah{1}<k<\ah{2}$, which means that this propagator is in the UV modes 
at $t=t_1$ while in the IR modes at $t=t_2$. 
Such propagators exist on the left-hand side of eq.~\eqref{funcdiv}, 
while they are absent on the right-hand side if we do not have the bilinear interaction. 
Furthermore, without the bilinear interaction, symmetric propagators of IR modes also vanish 
on the right-hand side because IR operators $\phi^<$ and $v^<$ are set to zero 
at the initial time $t=t_0$ and remain so. 
We show the explicit form of the bilinear interaction in sec.~\ref{frep} and prove eq.~\eqref{funcdiv} in sec.~\ref{jus}.

\subsection{Free propagators and interaction vertexes}\label{frep} 
We define the free part of the path integral in our formalism excluding the bilinear interaction by 
\begin{align}
Z_0[J=0]=\pathint\phi^<_{c,\Delta}\mathcal{D}v^<_{c,\Delta}~e^{iS^<_\mathrm{H,0}}\pathint\phi^>_{c,\Delta}\mathcal{D}v^>_{c,\Delta}~e^{iS^>_\mathrm{H,0}}\,.
\end{align}
In this expression, without interaction terms, 
it is obvious that UV modes and 
IR modes are treated independently.

The UV free propagators are given by a completely standard form as  
\begin{subequations}	
\label{uvprop1}
\begin{align}
G^{>ij}_{cc}(x,x')&\coloneqq\pathint v^>_{c,\Delta}\mathcal{D}\phi^>_{c,\Delta}~
{\phi_c^>}^i(x){\phi_c^>}^j(x')e^{iS^>_\mathrm{H,0}}=\frac{1}{2}\left<0\left|\left\{\phi^{>i}_\mathrm{I}(x),\phi^{>j}_\mathrm{I}(x')\right\}\right|0\right>\,,\\
G^{>ij}_{c\Delta}(x,x')&\coloneqq\pathint v^>_{c,\Delta}\mathcal{D}\phi^>_{c,\Delta}~
{\phi_c^>}^i(x){\phi_\Delta^>}^j(x')e^{iS^>_\mathrm{H,0}}=\Theta(t-t')\left[\phi^{>i}_\mathrm{I}(x),\phi^{>j}_\mathrm{I}(x')\right]\,,\\
G^{>ij}_{\Delta c}(x,x')&\coloneqq\pathint v^>_{c,\Delta}\mathcal{D}\phi^>_{c,\Delta}~
{\phi_\Delta^>}^i(x){\phi_c^>}^j(x')e^{iS^>_\mathrm{H,0}}=-\Theta(t'-t)\left[\phi^{>i}_\mathrm{I}(x),\phi^{>j}_\mathrm{I}(x')\right]=G^{>ij}_{c\Delta}(x',x)\,,\\
G^{>ij}_{\Delta\Delta}(x,x')&\coloneqq\pathint v^>_{c,\Delta}\mathcal{D}\phi^>_{c,\Delta}~
{\phi_\Delta^>}^i(x){\phi_\Delta^>}^j(x')e^{iS^>_\mathrm{H,0}}=0\,,
\end{align}
\end{subequations}
where $\phi^{1}\coloneqq\phi,\quad \phi^{2}\coloneqq v$.
From eq.~\eqref{uvprop1}, one can see that the $cc$ propagator is the symmetric propagator, the $c\Delta$ propagator is the retarded Green's function, and the $\Delta c$ propagator is the advanced Green's function. These propagators can be written more explicitly 
in terms of  
the mode functions as
\begin{subequations}
\begin{align}
G_{cc}^{>ij}(x,x')&=\modeint \ \Theta(k-\epsilon aH)
\,\Theta\left(k-\epsilon a(t')H\right)\, 
\mathrm{Re}\left[\Phi^i_k(t)\Phi^{*j}_k(t')\right]e^{i\vec k \cdot (\vec x - \vec{x}')}\,,\no \\
G_{c\Delta}^{>ij}(x,x')&=2i \,\Theta(t-t')\modeint \ \Theta(k-\epsilon aH)
\,\mathrm{Im}\left[ \Phi^i_k(t)\Phi^{*j}_k(t')\right]e^{i\vec k \cdot (\vec{x} - \vec{x}')}\,,\no
\end{align}
\end{subequations}
where
\begin{equation}
\Phi^1_k(t)\coloneqq\Phi_k(t),\ \Phi^2_k(t)\coloneqq\dot\Phi_k(t)\,. \label{modefndef}
\end{equation}

Next, we consider IR free propagators. 
Since the propagators including $v^<$ can be obtained 
by taking the derivative of $G^{<11}$ with respect to time, 
we concentrate on $G^{<11}$.\footnote{We should not take the time derivative of a step function $\Theta\left(\epsilon a(t')H-k\right)$ in a momentum integral: see eqs.~\eqref{irprop}.} 
As in the ordinary case, the retarded propagator obeys the equation
\begin{align*}
\Box_x\,G_{c\Delta}^{<11}(x,x')=\frac{i}{a^3}\delta(t-t')\modeint \ \Theta\left(\epsilon a(t')H-k\right)
\,e^{i\vec k \cdot (\vec{x} - \vec{x}')}\,, 
\end{align*}
and the boundary conditions are specified by 
$G_{c\Delta}^{<11}(x,x')=0$ for $t<t'$. 
Therefore, $G^{<ij}_{c\Delta}(x,x')$ can be written by 
means of the mode functions as
\begin{align*}
G_{c\Delta}^{<ij}(x,x')&=2i \,\Theta(t-t')\modeint \ \Theta\left(\epsilon a(t')H-k\right)
\,\mathrm{Im}\left[\Phi^i_k(t)\Phi^{*j}_k(t')\right]e^{i\vec k \cdot (\vec{x} - \vec{x}')}\,.
\end{align*}

On the other hand, the $cc$ propagator obeys
\begin{align*}
\Box_x\,G^{<11}_{cc}(x,x')=0\,, 
\end{align*}
with the boundary conditions that $G^{<11}_{cc}(x,x')=0$ at $t=t_0$, 
as we have assumed that there is no IR mode at $t=t_0$. 
Therefore, $G^{<11}_{cc}(x,x')$ vanishes identically. 
Since the $\Delta\Delta$ propagator $G^{<11}_{\Delta\Delta}(x,x')$ also obeys $\Box_x\,G^{<11}_{\Delta\Delta}(x,x')=0$, 
$G^{<ij}_{\Delta\Delta}(x,x')$ is identically zero for the same reason.

To summarize, the free propagators of IR modes are 
\begin{subequations}
\label{irprop}
\begin{align}
G_{cc}^{<ij}(x,x')&=0\,,\label{irprop1}\\
G_{c\Delta}^{<ij}(x,x')&=2i \Theta(t-t')\modeint \ \Theta\left(\epsilon a(t')H-k\right)\,\mathrm{Im}\left[\Phi^i_k(t)\Phi^{*j}_k(t')\right]e^{i\vec k \cdot (\vec{x} - \vec{x}')}\,,\label{irprop2}\\
G^{<ij}_{\Delta\Delta}(x,x')&=0\label{irprop3}\,.
\end{align}
\end{subequations}
The IR parts of $G^{ij}_{cc}(x,x')$, 
which exist in the ordinary QFT, 
are recovered by taking into account the bilinear interaction, 
as we shall prove in the next section.

Next we discuss the vertexes. 
In our formalism, there are two types of interaction terms. One is the 
usual self-interaction, and  the other is the bilinear interaction,  
which is peculiar to the present formulation.

An appropriate choice of the bilinear interaction 
$\tilde S_\mathrm{bilinear}$ is 
\begin{align}
&\tilde S_\mathrm{bilinear}[\irvd,\irphid,\uvvc,\uvphic]
\coloneqq-\int\mathrm{d}t\, a^3\modeint{}\delta(t-t_c(k))
\left[\irvd\left(\vec k,t\right)\uvphic\left(-\vec{k},t\right)-\irphid\left(\vec k,t\right)\uvvc\left(-\vec{k},t\right)\right]\,,
\label{biver}
\end{align}
where
\begin{align}
&\irvd\left(\vec k,t\right)=\int\mathrm{d}^3\! x\,\irvd\left(\vec x,t\right)e^{i\vec k\cdot\vec x}\,, \quad \irphid\left(\vec k,t\right)=\int\mathrm{d}^3 \! x\,\irphid\left(\vec x,t\right)e^{i\vec k\cdot\vec x}\,.
\end{align}
It should be notes that we cannot obtain the appropriate bilinear interaction \eqref{biver} by simply dividing the non-interacting part of Hamiltonian action $S_\mathrm{H,0}$ into UV parts and IR parts.
By dividing $S_\mathrm{H,0}$, we also obtain the bilinear interaction terms consisting of $c$-fields of IR modes and $\Delta$-fields of UV modes.
These vertexes denote the contribution coming from the flow from IR modes to UV mode. However, there is no such flow in de Sitter space, and hence we 
need to eliminate them by hand. 
In sec.~\ref{jus}, we show that indeed \eqref{biver} is an appropriate choice of bilinear interaction.

\begin{figure}[tbp]
 \centering
  \includegraphics[width=.7\textwidth,trim=80 610 100 130,clip]{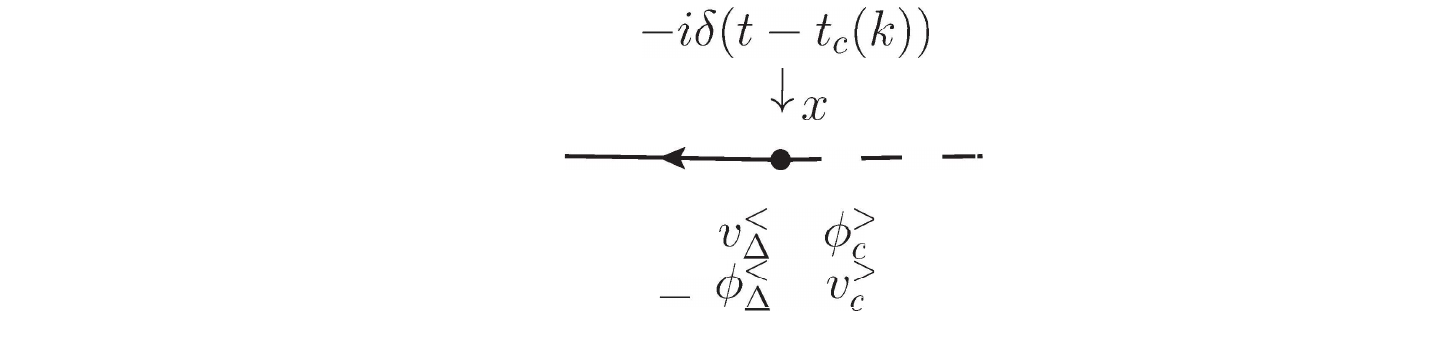}
 \caption{This diagram is the bilinear vertex. The corresponding vertex factor is also noted in this figure. Vertex integral is $\int\mathrm{d}t\,a^3$.}
 \label{bilinear5}
\end{figure}

We describe the bilinear interaction as a two-point vertex, 
which we call bilinear vertex, in the Feynman diagrams, as shown in fig.~\ref{bilinear5}. 
Owing to the factor $\delta(t-t_c(k))$ in~\eqref{biver}, 
the integrand is non-vanishing only for $k$ with $k=\epsilon aH$. 
As a Feynman rule, we assign 
$-i\delta(t-t_c(k))$ to the vertex $\irvd\left(\vec k,t\right)\uvphic\left(-\vec k,t\right)$, and $i\delta(t-t_c(k))$ to the vertex $\irphid\left(\vec k,t\right)v^>_c\left(-\vec k,t\right)$, and the vertex integral is taken over 
$\int\mathrm{d}t\,a^3$.

\begin{figure}[tbp]
\centering
  \includegraphics[width=.6\textwidth,trim=100 560 100 130,clip]{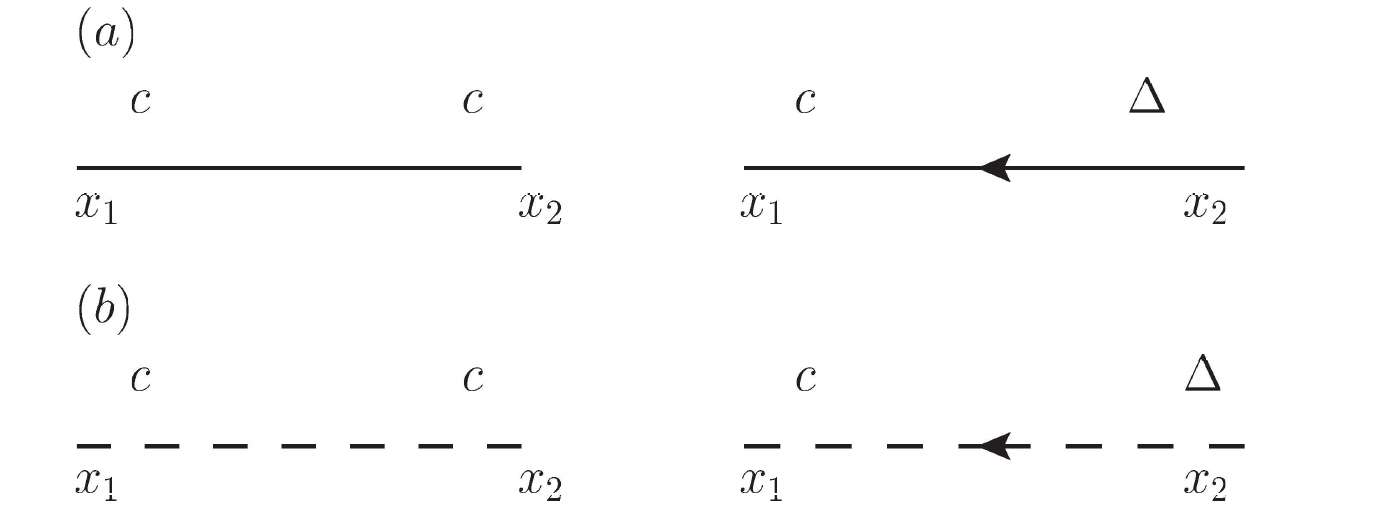}
\caption{Solid and dotted lines denote IR and UV propagators, respectively. 
Lines associated with an arrow correspond to $c\Delta$ propagators, while lines without an arrow correspond to $cc$ propagators. Therefore, the left-hand side of this figure shows $cc$ propagators, and propagators on the right hand side are $c\Delta$ propagators with $t_1\geq t_2$. }
 \label{uvcc}
\end{figure}

Self-interaction is described by $S_\mathrm{H,int}$, 
which is defined in eq.~\eqref{hamiact}. 
This vertex is treated just in the same way 
as in the usual perturbation theory.
Solid and dotted lines denote 
IR and UV propagators, respectively. 
Lines associated with an arrow correspond to $c\Delta$ propagators, 
while lines without an arrow correspond to $cc$ propagators, 
as shown in fig.~\ref{uvcc}. 
The arrow indicates the time direction. 
We explicitly associate $\phi$ or $v$ to the end of 
propagators when the distinction is necessary.

\subsection{The role of bilinear interaction and its justification}\label{jus}

Now, we prove that both sides of eq.~\eqref{funcdiv} are 
equivalent to each other. For this purpose, it is 
sufficient to show that the ordinary  
two point functions for a free field, 
$G_{IJ}(x,x')$, with $I,J$ being $c$ or $\Delta$, 
can be reproduced by 
taking into account 
the bilinear interaction.  
It is more convenient to 
focus on the Fourier component , 
$G_{IJ}\left(\vec k,t,\vec k',t'\right)$. 
We show the equivalence between $G_{IJ}\left(\vec k,t,\vec k',t'\right)$ 
and 
\begin{align}
\left<\phi^i_I\left(\vec k,t\right)\phi^j_J\left(\vec k', t'\right)\right>^{(0)}:=
\pathint\phi^<_{c,\Delta}\,\mathcal{D}v^<_{c,\Delta}~  
e^{iS^<_\mathrm{H,0}}\pathint\phi^>_{c,\Delta}
\,\mathcal{D}v^>_{c,\Delta}~e^{iS^>_\mathrm{H,0}}\,e^{i\tilde S_\mathrm{bilinear}}\,\phi^i_I\left(\vec k,t\right)\phi^j_J\left(\vec k', t'\right)
\,.
\label{justbili}
\end{align}

\begin{figure}[tbp]
 \centering
  \includegraphics[width=.7\textwidth,trim=80 600 100 130,clip]{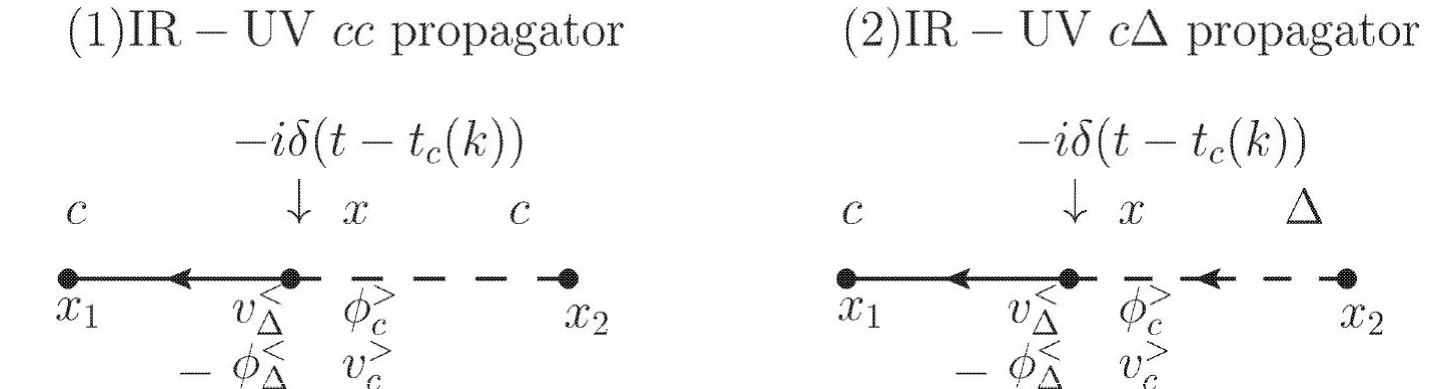}
 \caption{The $cc$ propagator with mode $k$ which satisfies $\ah{1}\leq k\leq\ah{2}$ is recovered  by connecting an IR $c\Delta$ propagator and a UV $cc$ propagator via the bilinear vertex.}
 \label{bilinear3}
\end{figure}

\begin{figure}[tbp]
 \centering
  \includegraphics[width=.43\textwidth,trim=120 590 120 100,clip]{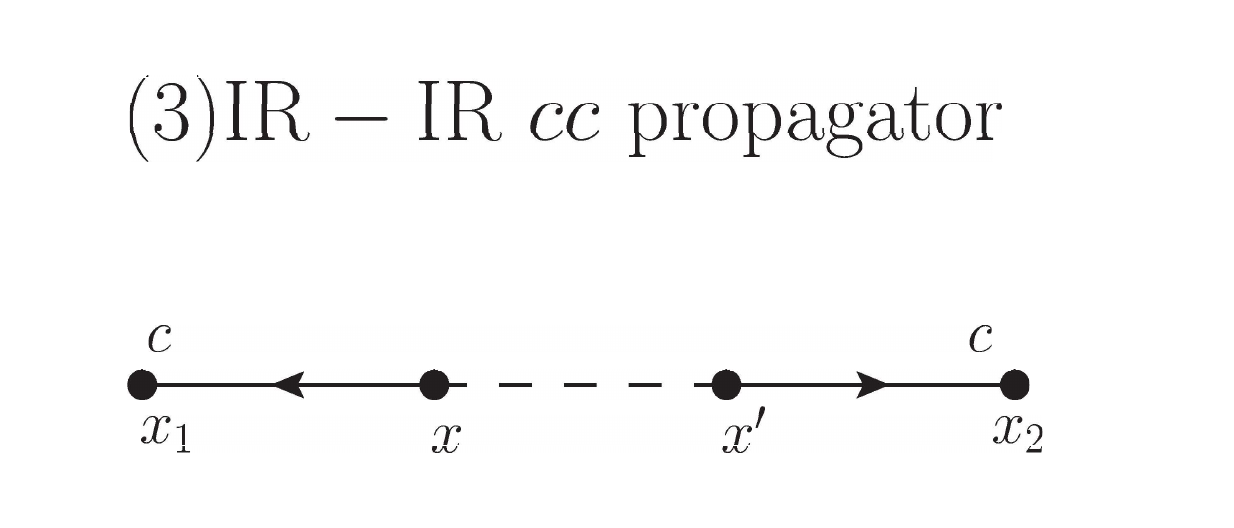}
 \caption{The propagator with mode $k$ which satisfies $k=\ah{1}=\ah{2}$  is recovered  by connecting two IR $c\Delta$ propagators and a UV $cc$ propagator via two bilinear vertexes.  }
 \label{bilinear4}
\end{figure}

Here, we stress that the bilinear interaction 
that we have introduced 
connects $\phi^{<i}_\Delta$ and $\phi^{>j}_c$. 
As 
the IR propagator vanishes for $G^{<ij}_{\Delta\Delta}$, 
the insertion of bilinear vertexes is relevant only when 
we consider the two-point functions that include 
$\phi^i_{c}\left(\vec k,t\right)$ with $t>t_c(k)$. 
In fact, the free propagators that we derived in the 
preceding subsection agree with the ordinary ones, 
except for  
$\left<\phi^i_c\left(\vec k,t\right)\phi^j_{c}\left(\vec k', t'\right)\right>^{(0)}$ 
and 
$\left<\phi^i_c\left(\vec k,t\right)\phi^j_{\Delta}\left(\vec k', t'\right)\right>^{(0)}$ 
with $t>t_c(k)$ and $t'<t_c(k')$,
and 
$\left<\phi^i_c\left(\vec k,t\right)\phi^j_c\left(\vec k', t'\right)\right>^{(0)}$ 
with $t>t_c(k)$ and $t'>t_c(k')$, which we call 0-th order 
IR-UV $cc$ and $c\Delta$, and 
IR-IR $cc$ two-point functions, respectively. Here, 0-th order means 
the order with respect to $\lambda$, and 0-th order two-point functions are distinguished from the free propagators. 
The Feynman diagrams that contribute 
to these 0-th order two-point functions from the bilinear interaction 
are presented in figs.~\ref{bilinear3} and \ref{bilinear4}. 
We discuss these diagrams in turn. 


\begin{enumerate}

\item IR-UV $cc$ two-point function\\
Assuming that $t_1>t_c(k)$ and $t_2<t_c(k')$, 
the contribution of the diagram (1) in 
fig.~\ref{bilinear3} corresponding to 
$\left<\phi^i_c\left(\vec k,t_1\right)\phi^j_{c}
\left(\vec k', t_2\right)\right>^{(0)}$, 
excluding the trivial factor $\delta^3\left(\vec k-\vec k'\right)$, 
is evaluated as 
\begin{align}
&\int\mathrm{d}t\,
 a^3\delta\left(t-t_c(k)\right)
 \frac{i}{2}\left[
 \left(\Phi^i_1\Phi^*-{\Phi^i_1}^*\Phi\right)
 \left(\dot\Phi{\Phi^j_2}^*+\dot\Phi^*\Phi^j_2\right)
  - \left(\Phi^i_1\dot\Phi^*-{\Phi^i_1}^*\dot\Phi\right)
 \left(\Phi{\Phi^j_2}^*+\Phi^*\Phi^j_2\right)\right]
\no\\ &
= \int\mathrm{d}t\,a^3
 \delta\left(t-t_c(k)\right)\frac{i}{2}
\left({\Phi^i_1}^*\Phi^j_2+\Phi^i_1{\Phi^j_2}^*\right)
\left(\dot\Phi\Phi^*-\Phi\dot\Phi^*\right)
= \frac{1}{2}
\left({\Phi^i_1}^*\Phi^j_2+\Phi^i_1{\Phi^j_2}^*\right)\,,
\label{recover1}
\end{align}
where $\Phi^{1}=\Phi$ , $\Phi^{2}=\dot\Phi$ as defined in eq.~\eqref{modefndef}. The subscripts 1, 2 associated with $\Phi^i$ 
label the time coordinate in its argument, and the subscript on 
the mode function $\Phi$ to specify the momentum $k$ 
is abbreviated, for notational simplicity.  
In the last equality, we used the normalization condition 
of mode functions~\eqref{eq:normalization}. 
The above result reproduces the ordinary expression for IR-UV $cc$ 
two-point function. 
\\

\item IR-UV $c\Delta$ two-point function\\
In a completely analogous manner, 
the contribution of the diagram (2) in 
fig.~\ref{bilinear3} correponding to 
$\left<\phi^i_c\left(\vec k,t_1\right)\phi^j_{\Delta}
\left(\vec k', t_2\right)\right>^{(0)}$ 
with $t_1>t_c(k)$ and $t_2<t_c(k')$ is evaluated as 
\begin{align}
&\int\mathrm{d}t\,a^3\delta\left(t-t_c(k)\right)
 i\left[\left(\Phi^i_1\Phi^*-{\Phi^i_1}^*\Phi\right)
 \left(\dot\Phi{\Phi^j_2}^*-\dot\Phi^*\Phi^j_2\right)
  - \left(\Phi^i_1\dot\Phi^*-{\Phi^i_1}^*\dot\Phi\right)
 \left(\Phi{\Phi^j_2}^*+\Phi^*\Phi^j_2\right)\right]
\no\\&
=\int\mathrm{d}t\, a^3\delta\left(t-t_c(k)\right)i
\left(\Phi^i_1{\Phi^j_2}^*-{\Phi^i_1}^*\Phi^j_2\right)
\left(\dot\Phi\Phi^*-\Phi\dot\Phi^*\right)
= \left(\Phi^i_1{\Phi^j_2}^*-{\Phi^i_1}^*\Phi^j_2\right)
\label{recover2}\,,
\end{align}
which reproduces the ordinary expression for IR-UV $c\Delta$ 
two-point function. 

\item 
IR-IR $cc$ two-point function\\
We evaluate the contribution of the diagram (3) in 
fig.~\ref{bilinear4} corresponding to 
$\left<\phi^i_c\left(\vec k,t_1\right)\phi^j_{c}
\left(\vec k', t_2\right)\right>^{(0)}$ 
with $t_1>t_c(k)$ and $t_2>t_c(k')$. 
We first perform the integration over $t'$,  
using the result of eq.~\eqref{recover1}.
Then, the remaining integration over $t$ is evaluated as 
\begin{align}
& \int\mathrm{d}t\, \delta\left(t-t_c(k)\right)
 \frac{-ia^3}{2}\biggl[
 \left(\Phi^i_1\Phi^*+{\Phi^i_1}^*\Phi\right)
 \left(\dot\Phi^*{\Phi^j_2}-\dot\Phi{\Phi^j_2}^*\right)
 -
 \left(\Phi^i_1\dot\Phi^*+{\Phi^i_1}^*\dot\Phi\right)
 \left(\Phi^*{\Phi^j_2}-\Phi{\Phi^j_2}^*\right)\biggl]
\no\\
&= \int\mathrm{d}t\, \delta\left(t-t_c(k)\right) 
 \frac{-ia^3}{2}\biggl[
 \left({\Phi^i_1}^*\Phi^j_2+\Phi^i_1{\Phi^j_2}^*\right)
 \left(\Phi\dot\Phi^*-\dot\Phi\Phi^*\right)
  \biggl]
=\frac{1}{2}({\Phi^i_1}^*\Phi^j_2+\Phi^i_1{\Phi^j_2}^*)
\label{recover3}\,,
\end{align}
which reproduces the ordinary expression for IR-IR $cc$ 
two-point function. 
\end{enumerate}

The above discussions show that all 0-th order two point functions 
are properly recovered by introducing the bilinear interaction 
$\tilde S_\mathrm{bilinear}$, and hence eq.~\eqref{justbili} is proved. 

\subsection{Effective EoM for IR modes}\label{efeom}

We have shown how to decompose the Schwinger-Keldysh path integral 
into the ones over IR modes and UV modes. 
Combining eqs.~\eqref{hamiact} and \eqref{funcdiv}, the 
generating functional for IR modes is written as 
\begin{align}
Z[J^<_{c,\Delta}]&=\pathint\phi^<_{c,\Delta}\mathcal{D}v^<_{c,\Delta}e^{iS^<_\mathrm{H}\left[\irvc,\irvd,\irphic,\irphid\right]+i\volint{x}\,\left(a^3{J^\phi_c}^<(x)\irphid(x)+{J^\phi_\Delta}^<\irphic(x)+\left(\phi^<\leftrightarrow v^<\right)\right)}e^{i\,\Gamma\left[\phi^<_{c,\Delta},v^<_\Delta\right]}\,,\label{IRgene}
\end{align}
with
\begin{align}
e^{i\Gamma}&=\pathint\phi^>_{c,\Delta}\mathcal{D}v^>_{c,\Delta}e^{iS^>_\mathrm{H}\left[v^>_c,v^>_\Delta,\uvphic,\uvphid\right]+i\tilde S_\mathrm{H,int}\left[\irphic,\irphid,\uvphic,\uvphid\right]+i\tilde S_\mathrm{bilinear}\left[\irvd,\irphid,v^>_c,\uvphic\right]}\label{influence}\,.
\end{align}
Here, $S^<_\mathrm{H}[\irvc,\irvd,\irphic,\irphid]\coloneqq
S_\mathrm{H}[\irvc,\irvd,\irphic,\irphid]$ 
is the part of the action purely composed of the IR modes, 
while 
$S^>_\mathrm{H}[v^>_c,v^>_\Delta,\uvphic,\uvphid]\coloneqq 
S_\mathrm{H}[v^>_c,v^>_\Delta,\uvphic,\uvphid]$ 
is that of the UV modes.  
$\tilde S_\mathrm{H,int}[\irphic,\irphid,\uvphic,\uvphid]
\coloneqq S_\mathrm{H,int}[\phi_c,\phi_\Delta]
-S^>_\mathrm{H,int}[\uvphic,\uvphid]-
S^<_\mathrm{H,int}[\irphic,\irphid]$ is the interaction 
part that depends on both IR and UV modes. 
The functional $\Gamma[\phi^<_{c,\Delta},v^<_\Delta]$ is 
obtained by integrating out UV modes and is called 
the influence functional, which contains 
all information of UV modes 
necessary to compute the generating functional for IR modes. 
$\Gamma$ can be calculated by evaluating 
all the connected diagrams, as usual. 
$\Gamma$ does not have the term which consists only of $\irphic$ fields,  
since $\Gamma$ must be odd under the overall exchange 
between (+) fields and (-) fields. 
Also, $\Gamma$ does not have $\irvc$ as it is not 
contained in the bare interaction vertexes.  

Now, we derive the effective EoM for IR fields for a given influence functional $\Gamma$. 
We decompose $S_{\rm H}^<$ and $\Gamma$ as
\begin{align}
 S_{\rm H,int}^<=S_{{\rm H(d)}}^<+S_{{\rm H(s)}}^<\,,\qquad
 \Gamma=\Gamma_{\rm (d)}+\Gamma_{\rm (s)}\,,
\end{align} 
where the first terms on the right hand side of the 
respective equations are the terms linear in $\irvd$ 
or $\irphid$, while the second terms express the other 
remaining pieces. 

Then, $\exp[iS^<_{\rm H(s)}+i\Gamma_{\rm (s)}]$ 
can be expressed by using the functional Fourier transformation as 
\begin{align}
e^{
 i\Gamma_{\rm (s)}\left[\irphic,\irphid,\irvd\right]
  +iS^<_{{\rm H(s)}} \left[\irphic,\irphid\right]}
=\int\mathrm{d}\xi_\phi\mathrm{d} \xi_v P\left[\xi_\phi,\xi_v,\irphic\right]
\exp\left[i\volint{x}\, a^3\xi_v\irphid
      -i\volint{x}\, a^3\xi_\phi\irvd\right]
      \,,\label{fourier}
\end{align}
where $\xi_\phi$ and $\xi_v$ are introduced as auxiliary fields. 
Then, the exponent on the right hand side 
becomes linear with respect to $\irphid$ or $\irvd$.
\footnote{In sec.~\ref{estimation}, we show that $iS^<_{{\rm H (s)}}$ term 
can be neglected in a good approximation since it 
turns out to be $\mathcal{O}(\epsilon)$. }
After this transformation, performing the path integral 
with respect to $\irphid$ and $\irvd$, we obtain the generating 
functional for IR modes as
\begin{align}
Z[J^<_{c,\Delta}]
&=\pathint\irphic\mathcal{D}\irvc \,
   e^{i\volint{x}\,a^3\left({J^\phi_\Delta}^<\irphic
    +{J^v_\Delta}^<\irvc\right)}\int\mathrm{d}
       \xi_\phi\mathrm{d}\xi_vP\left[\xi_\phi,\xi_v,\irphic\right]\no\\
 &\quad\times
    \delta\left(\dirphic-\irvc-\mu-\xi_\phi+{J_c^v}^<\right)
    \delta\left(\dirvc+3H\irvc+V_\mathrm{eff}'
       -\xi_v-{J_c^\phi}^<\right)\,,
\end{align}
where 
\begin{align}
 \mu=\frac{-1}{a^3}\frac{\delta}{\delta \irvd} 
     \left(S^<_{\rm H(d)}+\Gamma_{\rm (d)}\right)\,,
     \qquad
 V_\mathrm{eff}'=\frac{-1}{a^3}\frac{\delta}{\delta \irphid} 
     \left(S^<_{\rm H(d)}+\Gamma_{\rm (d)}\right)\,.
\end{align}
Therefore, the correlation functions of $c$-fields of IR modes can be evaluated by using $P[\xi_\phi,\xi_v,\irphic]$ as 
\begin{align}
&Z[0]^{-1}\left.\left[
 \frac{\delta}{ia^3(x_1)\delta{J^\phi_\Delta}^<(x_1)}
\cdots \frac{\delta}{ia^3(x_n)\delta{J^\phi_\Delta}^<(x_n)}
\frac{\delta}{ia^3(y_1)\delta{J^v_\Delta}^<(y_1)}
\cdots \frac{\delta}{ia^3(y_m)\delta{J^v_\Delta}^<(y_m)}
   \right]
Z[J^<_{c,\Delta}]\right|_{J^<_{c,\Delta}=0}
\no\\&
=\int\mathrm{d}\xi_\phi\, 
   \mathrm{d}\xi_v\, 
   P\left[\xi_\phi,\xi_v,\irphic\right]
    \left[\irphic(x_1)\cdots\irphic(x_n)\irvc(y_1)\cdots\irvc(y_m)
          \right]_{\dirphic-\irvc-\mu-\xi_\phi=0\,, 
      \ \dirvc+3H\irvc+V_{\mathrm{eff}}'(\irphic)-\xi_v=0}\,.\label{derieom}
\end{align}
Here, we use
\begin{align}
Z[0]&=\left.\int\mathrm{d}\xi_\phi\,
  \mathrm{d}\xi_v\,
   P\left[\xi_\phi,\xi_v,\irphic\right]
     \right|_{\dirphic-\irvc-\mu-\xi_\phi=0\,, 
       \ \dirvc+3H\irvc+V_{\mathrm{eff}}'(\irphic)-\xi_v=0}
\no\\
&=\left.\left[e^{i\Gamma_{\rm (s)}+iS^<_{{\rm H (s)}}}
          \right]_{
   \irphid,\irvd=0}\right|_{\dirphic-\irvc-\mu-\xi_\phi=0\,, 
   \ \dirvc+3H\irvc+V_{\mathrm{eff}}'(\irphic)-\xi_v=0}
=1\,.\label{normalization}
\end{align}
This equation shows that the weight function $P$ is automatically normalized as the consequence of the cancellation of the 
vacuum bubble diagrams in the in-in formalism.

Equation~\eqref{derieom} leads to the set of effective EoM for IR modes, 
\begin{subequations}
\label{generaleom}
\begin{align}
\dirphic&=\irvc+\mu(\irphic)+\xi_\phi \,,\\
\dirvc&=-3H\irvc-V_\mathrm{eff}'(\irphic)+\xi_v\,,\label{eom0}
\end{align}
\end{subequations}
which includes all quantum effects. 
Here, we neglect $\nabla^2\irphic/a^2$ term in eq.~\eqref{eom0} because this term vanishes in the limit $\epsilon\to 0$. 
Equation~\eqref{derieom} formally proves that the effective EoM~ \eqref{generaleom} can recover all correlation functions 
of $\irphic$ and $\irvc$ fields.
The auxiliary fields $\xi_\phi$ and $\xi_v$ behave as 
random variables, which follow the probability distribution 
$P[\xi_\phi,\xi_v,\irphic]$: 
\begin{align}
\left<
  \xi_v(x_1)\cdots\xi_v(x_n)\xi_\phi(y_1)\cdots\xi_\phi(y_m)
  \right>\coloneqq
\int\mathrm{d}\xi_\phi \mathrm{d}\xi_vP\left[\xi_\phi,\xi_v,\irphic\right]\xi_v(x_1)\cdots\xi_v(x_n)\xi_\phi(y_1)\cdots\xi_\phi(y_m)\,.
\end{align}
From this definition and eq.~\eqref{fourier}, one can see that noise correlations can be calculated for given $\Gamma$ as
\begin{align}
&\left<\xi_v(x_1)\cdots\xi_v(x_n)\xi_\phi(y_1)
    \cdots\xi_\phi(y_m)\right>\no\\
&=\left.\left[\left(
  \frac{\delta}{ia^3(x_1)\delta \irphid(x_1)}\right)\cdots\left(\frac{\delta}{ia^3(x_n)\delta \irphid(x_n)}\right)
   \left(\frac{-\delta}{ia^3(y_1)\delta \irvd(y_1)}\right)\cdots\left(\frac{-\delta}{ia^3(y_m)\delta \irvd(y_m)}\right)\right]
   e^{i\Gamma_{\rm (s)}+iS^<_{{\rm H (s)}}}\right|_{\irphid,\irvd=0}\,.\label{noisecalculation}
\end{align}

In order for the weight function $P[\xi_\phi,\xi_v,\irphic]$ 
to be interpreted as a probability distribution of 
$\xi_\phi$ and $\xi_v$,  
it should be real and positive.  We show that $P$ is real in general below.
Here, we leave to examine whether or not $P[\xi_\phi,\xi_v,\irphic]\geq 0$ 
holds in general, although 
the positivity of $P[\xi_\phi,\xi_v,\irphic]$ to NLO in 
a simple example becomes manifest in sec.~\ref{positivity}. 

Because flipping the sign of $\irphid$ and $\irvd$ is equivalent to take the complex conjugate
of the path integral $Z[0]$, and the functional measure $\mathcal{D}\irphid\mathcal{D}\irvd$ is invariant under these transformations, $i\Gamma_{\rm (s)}[\irphic,\irphid,\irvd]$ and $iS^<_{{\rm H (s)}}[\irphic,\irphid]$ must obey
\begin{equation}
i\Gamma_{\rm (s)}\left[\irphic,\irphid,\irvd\right]
+iS^<_{{\rm H (s)}}\left[\irphic,\irphid\right]
=\left(i\Gamma_{\rm (s)}\left[\irphic,-\irphid,-\irvd\right]
+iS^<_{{\rm H (s)}}\left[\irphic,-\irphid\right]\right)^*\,.\label{realproof1}
\end{equation}
\\
On the other hand, from eq.~\eqref{fourier}, $P$ and its complex conjugate $P^*$ can be written as
\begin{subequations}
\label{realproof2}
\begin{align}
P\left[\xi_\phi,\xi_v,\irphic\right]&=\int\mathrm{d}\irphid\int\mathrm{d}\irvd e^{i\Gamma_{\rm (s)}[\irphic,\irphid,\irvd]+iS^<_{{\rm H (s)}}[\irphic,\irphid]}e^{i\volint{x}a^3(\irvd(x)\xi_\phi(x)-\irphid(x)\xi_v(x))}\,,\\
\left(P\left[\xi_\phi,\xi_v,\irphic\right]\right)^*
&=\int\mathrm{d}\irphid\int\mathrm{d}\irvd e^{\left(i\Gamma_{\rm (s)}[\irphic,-\irphid,-\irvd]+iS^<_{{\rm H (s)}}[\irphic,-\irphid]\right)^*}e^{i\volint{x}a^3(\irvd(x)\xi_\phi(x)-\irphid(x)\xi_v(x))}\,.
\end{align}
\end{subequations}
Combining eqs.~\eqref{realproof1} and \eqref{realproof2}, $P$ turns out to be real,
\begin{equation}
P\left[\xi_\phi,\xi_v,\irphic\right]=\left(P\left[\xi_\phi,\xi_v,\irphic\right]\right)^*\in\mathbb{R}\,.
\end{equation}

\section{Systematic way of estimating the order of IR secular growth for each diagram and NLO stochastic dynamics}\label{systemNLO}
In order to derive an effective EoM for IR modes, one needs to calculate the influence functional $\exp\left[i\Gamma\right]$. In sec.~\ref{estimation}, we propose a systematic way of estimating the order of the IR secular growth for each diagram which constitutes the influence functional and $S^<_\mathrm{H,int}$.  In sec.~\ref{NLOderivation}, by using this proposed method, we explicitly derive the effective EoM for IR modes accurate enough to recover IR secular growth up to NLO in $\lambda\phi^4$ theory. In sec.~\ref{positivity}, we discuss the positivity of the weight function $P[\xi_\phi, \xi_v, \irphic]$. We show that the weight function is positive definite at least in the truncation maintaining up to NLO IR terms. In sec.~\ref{opor}, we comment on the restriction to the operator ordering of the correlation functions that we can calculate using our formalism.
\subsection{Systematic way of estimating the order of IR secular growth for each diagram}\label{estimation}
All the diagrams which constitute the influence functional are classified into two: diagrams which contribute to $V_\mathrm{eff}$ and diagrams which contribute to the noise correlation.
We call them $V_\mathrm{eff}$ diagrams and noise diagrams, respectively. 

\begin{figure}[tbp]
 \centering
  \includegraphics[width=.7\textwidth,trim=30 0 35 610,clip]{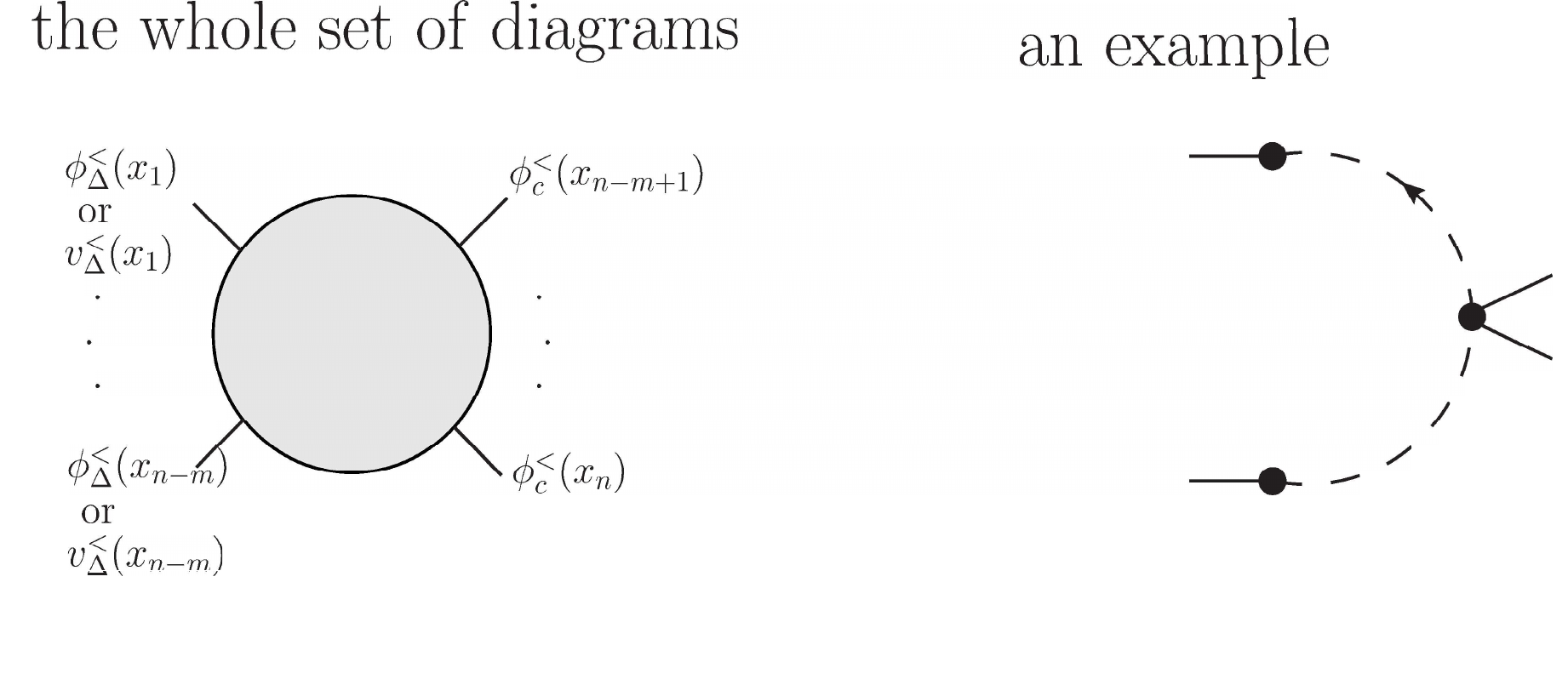}
 \caption{The diagram on the left-hand side represents the whole set of connected diagrams which constitute $i\Gamma$. Here, we adopt a rule that $\Delta$-fields of IR modes are written on the left and $c$-fields of IR modes on the right. The gray blob consists of diagrams connected only by UV propagators. A concrete example is shown on the right-hand side. }
 \label{IRestimation}
\end{figure}
 All the connected diagrams are represented as a whole by the left diagram in fig.~\ref{IRestimation}. $\Delta$-fields of IR modes are written on the left and $c$-fields of IR modes on the right. The gray blob consists of diagrams connected only by UV propagators. A concrete example is given by the right diagram in fig.~\ref{IRestimation}. Since all the vertexes are connected by UV propagators, it is expected that the influence functional $i\Gamma$ is approximately local. 
 That is, if one consider a certain diagram with $n$ external vertexes $x_1\cdots x_n$, the support of the diagram will be approximately given by $\vec{x}_i\sim\vec{x}_j$ and $t_i\sim t_j$ for arbitrary pairs of $(i,j),\ i,j=1,\cdots, \, n$,  and outside this region the value of diagram will decay exponentially. Here, $\vec{x}_i\sim\vec{x}_j$ and $t_i\sim t_j$ mean that $|\vec{x}_i-\vec{x}_j|\lesssim1/\ah{i}\sim1/\ah{j}$ and $|t_i-t_j|\lesssim\ln(1/\epsilon)/H$, respectively. In this section, we assume this approximate locality is satisfied by all diagrams: see appendix~\ref{colliapp} for the discussion about the validity of this assumption. Under this assumption, we develop a systematic way  to estimate the order of IR secular growth for each diagram, or equivalently counting the number of the factor of $\ln (a/a_0)$. 

Since the time evolution of correlation functions is determined by eq.~\eqref{generaleom}, one can evaluate them in terms of the time integration of the noise correlations for given initial conditions. Then, the factor of $\ln (a/a_0)\propto t-t_0$ comes from the integration over time $\int\mathrm{d}t'$ from $t_0$ to $t$. Therefore, in order to estimate the order of IR secular growth for each diagram, we need to count the number of the time integrals included in the diagram whose effective integration range is comparable to the whole range of time that we are concerned with. At first, we express $\irphic$ in terms of noise fields. Both $\xi_\phi$ and $\xi_v$ can be treated simultaneously by the unified noise $\tilde \xi$, which is defined by
\begin{equation}
\tilde\xi\coloneqq 3H\xi_\phi+\dot\xi_\phi+\xi_v\,.\label{unifiednoise}
\end{equation}
By using the unified noise $\tilde\xi$, eq.~\eqref{generaleom} is rewritten as 
\begin{equation}
\ddirphic+3H\dirphic=-V'_\mathrm{eff}(\irphic)+\dot\mu(\irphic)+3H\mu(\irphic)+\tilde\xi\,. \label{unifiedeom}
\end{equation}
From this equation, $\irphic(T)$ can be expressed by using $\tilde\xi(t)$ formally as
\begin{align}
\irphic(T)&=i\int^{T}_{t_0}\mathrm{d}t \, a^3G(T,t)\left[-V'_\mathrm{eff}+\dot\mu+3H\mu+\tilde\xi(t)\right]\,,\label{formalsol}\\
G(T,t)&\coloneqq\frac{i}{3H}\left(\frac{1}{a^3(T)}-\frac{1}{{a}^3}\right)\Theta(T-t)\,, \quad -\left(\der^2_T+3H\der_T\right)G(T,t)=\frac{i}{a^3(T)}\delta(T-t)\,, \label{Greenfn}
\end{align}
which allows us to express $\irphic(T)$ in terms of $\tilde\xi$ iteratively expanding with respect to the coupling constant. We will take into account the self-interaction terms later and concentrate on 0-th order with respect to the coupling constant now. Then, the number of time integrals to be assigned to $\irphic$ is counted as $1-(1/2)=1/2$ because the noise correlation $\left<\tilde\xi\tilde\xi\right>$ is given by eq.~\eqref{LOstoc} at 0-th order. Next, we establish how to assign the number of time integrals to each diagram with $\Delta$-fields of IR modes. When a $\Delta$-field of IR modes is contracted with $\irphic$, one gets retarded Greens functions $G^{<1i}_{c\Delta}$. Although the number of time integrals attributed to $\irphic$ is counted as $1/2$, retarded Green's functions $G^{<1i}_{c\Delta}$ themselves do not grow at a late time, which implies that the number of time integrals attributed to $\irphic$ is canceled by $\irphid$, and hence the number of time integral to be assigned to a $\Delta$-field of IR modes is counted as $-1/2$ at 0-th order. Finally, from vertex integrals included in the gray blob in fig.~\ref{IRestimation}, only one time integral remains unconstrained because of the approximate locality of $i\Gamma$. From now on, we refer to this time integral as the time integral of the vertex.

For example, let us consider $\lambda\phi^4$ theory. From the above discussion, the number of time integrals attributed to the diagram with $n-m$ $\Delta$-fields and $m$ $c$-fields of IR modes, which is shown in fig.~\ref{IRestimation}, is $1+(m/2)-\left((n-m)/2\right)$.
Suppose that this diagram is $\mathcal{O}\left(\lambda^l\right)$.
Then, this diagram contributes to IR secular growth terms in correlation functions to  
\begin{equation}
\mathrm{N}^{2l-\left[1+(m/2)-\left((n-m)/2\right)\right]}\mathrm{LO}=\mathrm{N}^{(n/2)+2l-(m+1)}\mathrm{LO}\,,\label{IRstrength}
\end{equation}
because the $l$-th order vertex in $\lambda$ which consists only of LO vertex $\lambda\irphic^3\irphid$ has $(3l/2)-(l/2)+l=2l$ time integrals, and the difference between $2l$ and $1+(m/2)-\left((n-m)/2\right)$ is $(n/2)+2l-(m+1)$. 
We can gather the diagrams necessary to derive the stochastic formalism described by eq.~\eqref{LOstoc}, which is valid up to LO, by using eq.~\eqref{IRstrength}: see appendix~\ref{exercise}.

\begin{figure}[tbp]
 \centering
  \includegraphics[width=.7\textwidth,trim=50 650 50 130,clip]{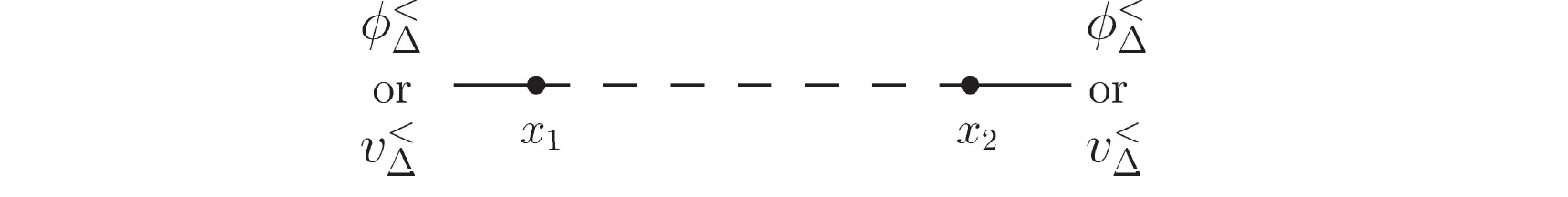}
\caption{These noise diagrams contribute to the LO noise correlation which is described by eq.~\eqref{LOstoc}. More precisely, the diagram shown here without $\irphid$ fields contribute to the LO noise correlation, and the remaining diagrams are $\mathcal{O}(\epsilon)$ suppressed.}
 \label{snownoise1}
\end{figure}

\begin{figure}[tbp]
 \centering
  \includegraphics[width=.25\textwidth,trim=180 580 170 120,clip]{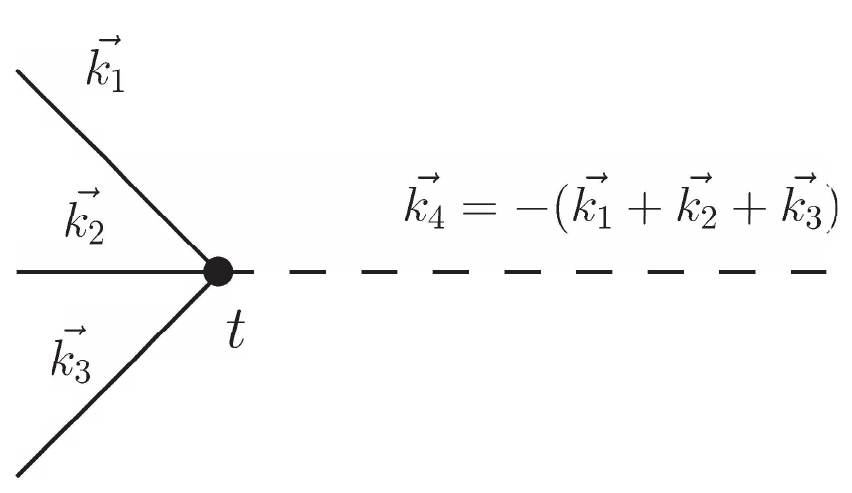}
 \caption{Three IR propagators and one UV propagator are combined via this vertex. }
 \label{birdfootvertex} 
\end{figure}

We have established the counting rule, but the vertexes with 
one UV leg suffer from further suppression. We assign momenta as shown in fig.~\ref{birdfootvertex}. 
Although three IR propagators are attached to this vertex, 
the range of the time integral is suppressed unless two of the momenta of IR propagators are around $\epsilon aH$ due to the momentum conservation. If we assume that an inequality $k_1\sim k_2\lesssim\delta\epsilon aH\ll k_3\sim\epsilon aH$ with $\delta\ll1$ and $\ln(1/\delta)\ll\ln(a/a_0)$ holds,
the integration range of the integral over $\vec{k}_3$ is restricted to very narrow region:
\begin{align*}
\epsilon aH\left(1-\frac{\bigl|\vec{k}_1+\vec{k}_2\bigr|}{\epsilon aH}\right) \lesssim k_3\leq \epsilon aH\,,
\end{align*}
where higher order terms with respect to $\bigl|\vec{k}_1+\vec{k}_2\bigr|/k_3$ are neglected. When one evaluates the contribution from this vertex to correlation functions, this restriction of the integration range implies that the remaining integrand after performing the integration over $\vec{k}_3$ is suppressed by factor $\bigl|\vec{k}_1+\vec{k}_2\bigr|/\epsilon a H$, compared to the case without this restriction. 
Therefore, contribution from this vertex is suppressed by factor $\mathcal{O}(\delta)$ when two of three momenta are deep IR modes, which means that one has to put two momenta around $\epsilon aH$ in evaluating the contribution from this vertex to correlation functions. For this reason, one needs to change the counting rule mentioned above in evaluating the diagram containing the vertexes with one UV leg. 
 From now on, we refer to a field whose assigned momentum is restricted to $\delta\epsilon a H\lesssim k<\epsilon aH$ as a restricted field. 
In estimating the order of IR secular growth, IR propagators including restricted fields can be replaced by UV propagators with mode $\vec k$ satisfying $k\sim \epsilon aH$. This is because the contribution from IR propagators including restricted $\irphic$ are simply obtained by changing the lower bound of momentum integration of UV propagators from $\epsilon aH$ to $\delta\epsilon aH$ with $\delta\ll1$ and $\ln(1/\delta)\ll\ln(a/a_0)$, and this replacement does not change the order of magnitude. Therefore, in evaluating the order of IR secular growth, one can replace vertexes $\lambda\irphid\irphic^2\uvphic$ and $\lambda\irphic^3\uvphid$ by $\lambda\irphic\uvphid\uvphic^2$ or $\lambda\irphid{\phi^>}^3$, and $\lambda\irphic\uvphid\uvphic^2$, respectively.

\begin{figure}[tbp]
 \centering
  \includegraphics[width=.41\textwidth,trim=170 650 180 131,clip]{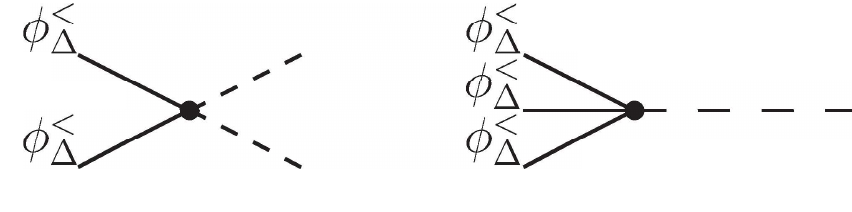}
 \caption{Examples of vertexes which have several $\Delta$-fields of IR modes.}
 \label{deltas}
\end{figure}

There are also vertexes which can only contribute to $\mathcal{O}(\epsilon)$ terms in correlation functions and hence are irrelevant in our analysis. The Fourier components of IR propagators $\tilde G^{<ij}_{cc}(t,t';k)$ and $\tilde G^{<ij}_{c\Delta}(t,t';k)$ satisfy the following relations: 
\[
\frac{\tilde G^{<i1}_{c\Delta}(t,t';k)}{\tilde G^{<i1}_{cc}(t,t';k)}\sim\begin{cases}
\mathcal{O}\left(\epsilon^3\right) & i=1~, \\
\mathcal{O}(\epsilon) & i=2~.
\end{cases}
\]
Therefore, in evaluating correlation functions which consist of $\irphic$ and $\irvc$, we can neglect diagrams which include vertexes with more than one $\irphid$. 
For example, we can neglect $S^<_{{\rm H (s)}}$ and the vertexes shown in fig.~\ref{deltas}.

\subsection{NLO stochastic dynamics}\label{NLOderivation}
\begin{figure}[tbp]
 \centering
  \includegraphics[width=.77\textwidth,trim=30 10 30 400,clip]{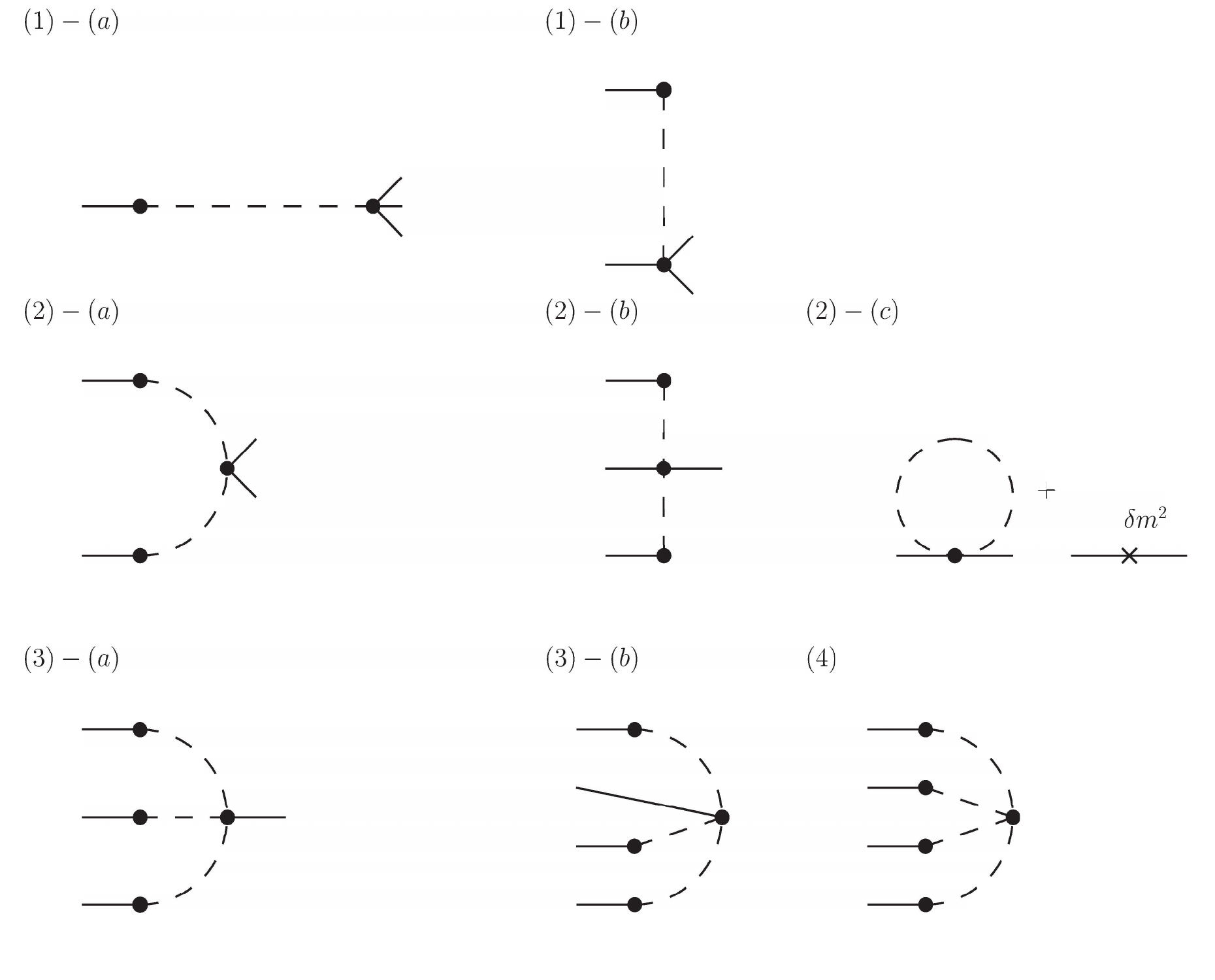}
 \caption{The diagrams which are first order with respect to $\lambda$ are shown. Here we adopt the same rule as fig.~\ref{IRestimation}: $\Delta$-fields of IR modes are written on the left and $c$-fields of IR modes on the right. Only in this figure, dotted lines denote both of UV $cc$ propagators and UV $c\Delta$ propagators simultaneously because it is irrelevant to distinguish between UV $cc$ propagators and UV $c\Delta$ propagators in estimating the order of IR secular growth by using eq.~\eqref{IRstrength}.}
\label{NLOcan2}
\end{figure}

\begin{figure}[tbp]
 \centering
  \includegraphics[width=.92\textwidth,trim=30 10 30 625,clip]{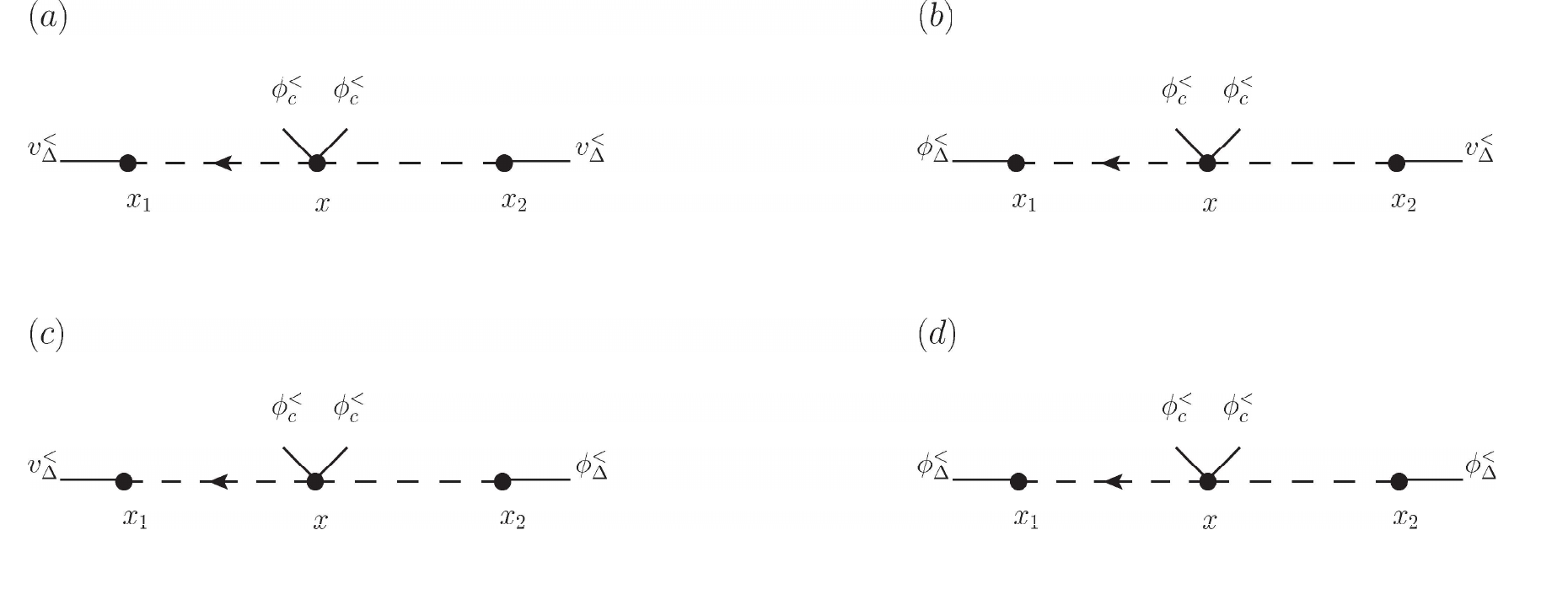}
 \caption{These diagrams can contribute to NLO IR secular growth according to the estimation by using eq.~\eqref{IRstrength}.}
 \label{Next-1}
\end{figure}

Now, let's move on to derive an effective EoM which can describe IR secular growth up to NLO. From now on, we refer to this effective EoM  as NLO stochastic EoM. In order to derive NLO  stochastic EoM, let us consider the number of time integrals to be assigned to the vertexes with UV legs which constitutes the diagrams shown in fig.~\ref{IRestimation}. From the discussion in sec.~\ref{estimation}, the number of time integrals attributed to a vertex with $p$ UV legs and $q$ $\irphid$ fields is counted as $-(p/2)-q$ for $p\geq2$ compared to the leading order vertex $\lambda\irphid\irphic^3$. We need to take care when $p=1$, as we discussed at the end of sec.~\ref{estimation}:  
$p=1$ vertex is equivalent to $p=3$ vertex regarding this counting because of the presence of restricted fields. Therefore, maximum value attributed to a vertex with UV legs is $-1$, which is realized for $(p,q)=(2,0)$, and hence the order of IR secular growth decreases as the number of these vertexes increases. Then, one may think that the most dominant $l$-th order diagram at a late time would be given by the diagram which only consists of $\lambda\irphic^2\uvphic\uvphid$ vertex, which would contribute to $\mathrm{N}^{l-1}\mathrm{LO}$ IR secular growth  where $-1$ coming from the time integral of the vertex. However, all the diagrams must have at least one $\Delta$-field of IR modes from the requirement of causality, and this cancel the time integral of the vertex. Thus, $l$-th order diagrams can contribute to $\mathrm{N}^l\mathrm{LO}$ at most,
 and we need to consider only first order diagrams with respect to $\lambda$ to derive NLO stochastic EoM. These diagrams are listed up in fig.~\ref{NLOcan2}. Note that only in fig.~\ref{NLOcan2}, dotted lines denote either UV $cc$ propagators or UV $c\Delta$ propagators simultaneously because it is irrelevant to distinguish between UV $cc$ propagators and UV $c\Delta$ propagators in estimating the order of IR  secular growth by using eq.~\eqref{IRstrength}.
Then, one can easily see that diagrams (2)-(a) and (2)-(c) contribute to NLO IR secular growth. 
However, since the value of UV loop integral in diagram (2)-(c) is time independent, the contribution from diagram (2)-(c) can be exactly canceled by  an appropriate choice of the finite term in the mass renormalization $\delta m^2$, and we choose so in our study. 
Thus, what we need to calculate is only diagram (2)-(a).  
Depending on the choice of  $\Delta$-fields of IR modes at endpoints, there are 4 diagrams(a)-(d): see fig.~\ref{Next-1}. 

The contribution from diagram (a) to $i\Gamma$, which we denote as $i\Gamma_\mathrm{(a)}$, is evaluated as 
\begin{align}
i\Gamma_\mathrm{(a)}&=\volint{x_1}\,{a^3(t_1)}\irvd(x_1)\volint{x_2}\,{a^3(t_2)}\irvd(x_2)N_\mathrm{(a)}(x_1,x_2)\,,\no\\
N_\mathrm{(a)}(x_1,x_2)&\sim\int\mathrm{d}^3 \! x\int^{t_m}_{t_0}\mathrm{d}t\,a^3\irphic^2(\vec x,t)\modeint{1}\modeint{2}\no\\
&\quad \times\tilde G^{>11}_{c\Delta}(t_1,t;k_1)\tilde G^{>11}_{cc}(t_1,t;k_2)e^{i\vec{k}_1\cdot(\vec{x}_1-\vec x)}e^{i\vec{k}_2\cdot(\vec{x}_2-\vec x)}\delta(t_1-t_c(k_1))\delta(t_2-t_c(k_2))\,, \label{effaction1}\\
G^{>ij}(x,x')&=\modeint{}\tilde G^{>ij}(t,t';k)e^{-i\vec k\cdot(\vec x-\vec{x}')}\,.\no
\end{align}
Here, $t_m\coloneqq\mathrm{min}(t_1,t_2)$ and we omit a proportionality coefficient in the 2nd line. Up to NLO, we only need to extract the most dominant IR secular growth from $\irphic(\vec x,t)$ in the integrand because these noise diagrams contribute to NLO at most. In this case, when the range of the vertex integral $\int\mathrm{d}^4\! x$ is restricted to the coarse-graining scale around a certain space-time point $X=(\vec X,T)$, we can replace $\irphic(\vec x,t)$ by $\irphic(\vec X,T)$. This is because the most dominant part of $\irphic(\vec x,t)$ at a late time is identical to that of $\irphic(\vec X,T)$ if $\vec x\sim\vec X$ and $t\sim T$.\footnote{This statement can be shown as follows. First, let us consider the time dependence of $\irphic$. From eqs.~\eqref{formalsol} and \eqref{Greenfn}, the relationship between $\irphic(\vec x,t)$ and $\irphic(\vec x,T)$ is given by
\begin{align*}
\irphic(\vec x,t)&=\irphic(\vec x,T)-i\int^{T}_t\mathrm{d}t'\, {a}^3(t') G(T,t')\left[-V'_\mathrm{eff}+\dot\mu+3H\mu+\tilde\xi(t')\right]\,.
\end{align*}
If $T-t\ll T-t_0$ is satisfied, it is obvious that only $\irphic(\vec x,T)$ is the most dominant term on the right-hand side. Next, let us consider the spatial dependence of $\irphic(\vec x, t)$. The most dominant part of $\irphic(\vec x,t)$ at a late time appears only when the momentum $k$ which is assigned to $\irphic$ is much less than $\epsilon aH$. And if both $k\ll\epsilon aH$ and $|\vec{x}_1-\vec x|,\ |\vec{x}_2-\vec x|\lesssim1/(\epsilon aH)$ are satisfied, we can replace $e^{i\vec k\cdot\vec x}$ by $e^{i\vec k\cdot\vec{x}_1}$ or $e^{i\vec k\cdot\vec{x}_2}$ in a good approximation. Therefore if $|\vec{x}_1-\vec x|,\ |\vec{x}_2-\vec x|\lesssim1/(\epsilon aH)$ are satisfied, we can use the equality $\irphic(\vec x,t)\simeq\irphic(\vec{x}_1,t)\simeq\irphic(\vec{x}_2,t)$ regarding the most dominant part of $\irphic(\vec x,t)$ at a late time.}
Now, we prove that the range of vertex integral $\int\mathrm{d}^4\! x$ in eq.~\eqref{effaction1}  is restricted to the coarse-graining scale around a space-time point $x_1\sim x_2\sim X$. First, we consider the spatial integral. From the angular integrals of $\vec k_1$ and $\vec k_2$ in eq.~\eqref{effaction1}, $j_0(k_1|\vec x_1-\vec x|)$ and $j_0(k_2|\vec x_2-\vec x|)$ appear in the integrand, which lead to exponential decay for $|\vec x_1-\vec x|,\ |\vec x_2-\vec x|\gg1/(\epsilon aH)$ after time coarse-graining. 
Next, we show that the integrand in eq.~\eqref{effaction1} is exponentially suppressed for $t_m-t\gg1/H$. For $t_m-t\gg1/H$, the momenta $k_3$ and $k_4$ which are assigned to $\irphic^2(\vec x,t)$ are much less than $k_1$ and $k_2$ because the upper bound for $k_3$ and $k_4$ is given by $\epsilon aH$. In this case, we can set $\vec k_1+\vec k_2=\vec 0$, which results in
\begin{align*}
N_\mathrm{(a)}(x_1,x_2)
\sim \int^{t_m}_{t_0}\mathrm{d}t\, \tilde\irphic(\vec k_3,t)\tilde\irphic(-\vec k_3,t)\cdot H^2\delta(t_1-t_2)&j_0(\ah{1}|\vec x_1-\vec x_2|)\\
&\times\left(\frac{a}{a(t_1)}\right)^3\frac{1}{\epsilon^3}\left(f_I(\Delta t_1)+\frac{\epsilon^3}{3}f_R(\Delta t_1)\right)f_R(\Delta t_1)\,,
\end{align*} 
where $f_R$ and $f_I$ are defined in eq.~\eqref{defoff}, $\Delta t_1\coloneqq t_1-t$, and $j_0(z)\coloneqq \sin z/z$. From the above equation and eq.~\eqref{defoff}, one can see that the integrand in eq.~\eqref{effaction1} decays exponentially for $t_m-t\gg1/H$. 
It is straightforward to apply the above discussion to other diagrams in fig.~\ref{Next-1}.

Now it is justified to replace $\irphic(\vec x,t)$ by $\irphic(\vec x_1,t_1)$ in eq.~\eqref{effaction1}, and we get 
\begin{align}
N_\mathrm{(a)}(x_1,x_2)
&\simeq-\frac{H\lambda}{24\pi^2}\irphic^2(x_1)
\left(-\ln\frac{1}{\epsilon}+\ln2+\gamma-2+\mathcal{O}\left(\epsilon^2\right)\right)\delta(t_1-t_2)j_0(\ah{1}|\vec x_1-\vec x_2|)\,.\label{NLOa}
\end{align}
Note that the approximation adopted in the first line holds up to NLO.\\
Similarly, diagram (b) is calculated as follows:
\begin{align}
i\Gamma_\mathrm{(b)}&=\volint{x_1}\, a^3(t_1)\irphid(x_1)\volint{x_2}\,a^3(t_2)\irvd(x_2)N_\mathrm{(b)}(x_1,x_2)\,,\no\\
N_\mathrm{(b)}(x_1,x_2)
&\simeq\frac{-\lambda H^2\irphic^2(x_1)}{24\pi^2}\left(1+\mathcal{O}\left(\epsilon^2\right)\right)\delta(t_1-t_2)j_0(\ah1|\vec x_1-\vec x_2|)\,.\label{NLOb}
\end{align}
Similarly we can evaluate diagrams (c) and (d), and it turns out that they are $\mathcal{O}(\epsilon)$ suppressed.

Therefore, combining eqs.~\eqref{generaleom}, \eqref{noisecalculation}, \eqref{NLOa}, and \eqref{NLOb}, we obtain a set of effective EoM for IR modes which can describe IR secular growth up to NLO:
\begin{subequations}
\label{nlosto}
\begin{align}
\dirphic&=\irvc+\xi_\phi\,,\quad \dirvc=-3H\irvc-\frac{\lambda\irphic^3}{6}+\xi_v\,,\label{langevin}\\
\left<\xi_{\phi}(x_1)\xi_{\phi}(x_2)\right>&=\frac{H}{4\pi^2}\left[H^2+\lambda l(\epsilon)\irphic^2(x_1)\right]\delta(t_1-t_2)j_0(\epsilon a(t_1)H|\vec x_1-\vec x_2|)\,,  \label{phiphinoise} \\
\left<\xi_{\phi}(x_1)\xi_v(x_2)\right>&=-\frac{H^2\lambda}{24\pi^2}\irphic^2(x_1)\delta(t_1-t_2)j_0(\epsilon a(t_1)H|\vec x_1-\vec x_2|)\,,\label{vphinoise} \\ 
\left<\xi_v(x_1)\xi_v(x_2)\right>&=\mathcal{O}\left(\epsilon^2\right)\simeq0\,, \quad l(\epsilon)\coloneqq-\frac{1}{3}\left[\ln\frac{1}{\epsilon}-\ln2-\gamma+2\right]\,.\label{vvnoise}
\end{align}
\end{subequations}
Since the time derivative of $\xi_v$ cannot contribute to NLO, we can further simplify the above stochastic EoM by redefining the $\irvc$, up to NLO, as 
\begin{subequations}
\label{nlosto2}
\begin{align}
\dirphic&=\irvc+\xi'_\phi\,,\quad \dirvc=-3H\irvc-\frac{\lambda\irphic^3}{6}\,,\\
\left<\xi'_{\phi}(x_1)\xi'_{\phi}(x_2)\right>&=\frac{H}{4\pi^2}\left[H^2+\lambda\irphic^2(x_1)\left(l(\epsilon)-\frac{1}{9}\right)\right]\delta(t_1-t_2)j_0(\epsilon a(t_1)H|\vec x_1 -\vec x_2|)\no\\
&\coloneqq N\delta(t_1-t_2)j_0(\epsilon a(t_1)H|\vec x_1-\vec x_2|)\,.
\end{align}
\end{subequations}
Here, $\xi'_\phi$ is related to $\xi_\phi$ and $\xi_v$ as 
\begin{align*}
\xi'_\phi=\xi_\phi+\frac{1}{3H}\xi_v\,.
\end{align*} 
Let us compare this result with LO stochastic EoM described by eq.~\eqref{LOstoc}. LO stochastic dynamics is just a Brownian motion under an external force with field-\emph{independent}  Gaussian white random noise. On the other hand, NLO stochastic dynamics is a Brownian motion under the same external force as in the LO case with field-\emph{dependent} Gaussian white random noise. Another difference between LO stochastic EoM and NLO stochastic EoM is that second order time derivative term of $\irphic$ appears in eq.~\eqref{nlosto2}, although this difference hardly changes the dynamics qualitatively. The properties of time and spatial correlations are unchanged, and the system has no hysteresis. 

In order to clarify whether or not IR secular growth can be interpreted as an increase of the statistical ensemble average of the variance at least up to NLO, it is needed to show the positivity of the corresponding weight function of the noise. We will show this positivity in the next sec.~\ref{positivity}.

It should be noted that the discussion parallel to the above applies to the theory with a general power-law potential $V(\phi)=g\phi^n$. NLO stochastic EoM for this theory is obtained by simply replacing $\lambda\irphic^3/6$ and $\lambda\irphic^2$ by $V'(\irphic)$ and $2V''(\irphic)$ in eq.~\eqref{nlosto}, respectively. NLO correction to LO stochastic EoM is simply caused by UV modes acquiring effective mass.

\subsection{Positivity of the weight function}\label{positivity}
In this section, we discuss the positivity of the weight function $P[\xi_\phi, \xi_v, \irphic]$ 
which is defined in eq.~\eqref{fourier}, especially the positivity of the weight function of the noise which contributes to NLO IR secular growth at least. 
Since NLO stochastic dynamics is described by the Gaussian noise $\xi'_\phi$, the corresponding weight function is positive unless the amplitude of the noise becomes negative. 
As you can see from the expression eqs.~\eqref{nlosto2}, correction terms to the noise amplitude relative to leading order terms depend on $\lambda$ and $\irphic/H$. In a region where a potential $V(\phi)$ is sufficiently flat and the absolute value of $\lambda(\irphic/H)^n$ for $n=1,2,3$ is small, sub-leading terms for each noise correlation are small.\footnote{When $\phi$ is an inflaton, this condition corresponds to the slow-roll condition.} Outside this region this condition does not hold, and UV modes and IR modes are strongly interacting with each other, which implies that we cannot handle this theory perturbatively.
Thus, as far as UV modes and IR modes are weakly interacting with each other, the positivity of the weight function is ensured at least up to NLO. Furthermore, NLO stochastic dynamics is Markovian, and the initial time $t_0$ can be smoothly sent to the past infinity. Therefore, NLO stochastic dynamics can be seen as a classical stochastic process. Note that it is straightforward to extend the discussion here to the theory with a power-law potential $V(\phi)=g\phi^n$. 

However, it seems difficult to show the positivity of $P[\xi_\phi, \xi_v, \irphic]$ beyond NLO, because $P$ is obtained by Fourier transforming a non-Gaussian functional. 
Further investigation is needed. We will study this aspect further in our future work. 

\subsection{Operator ordering}\label{opor}

In sec.~\ref{NLOderivation}, we derived the effective EoM of $\irphic$ and $\irvc$ fields. However, the operator orderings of correlation functions which can be calculated by using our extended stochastic formalism are uniquely fixed: we can calculate only correlation functions consisting of $\irphic$ and $\irvc$ fields in our formalism, as we mentioned in sec.~\ref{efeom}. 
In order to obtain correlation functions with arbitrary operator orderings, we also need to know correlation functions which include $\Delta$-fields. However, such correlation functions are $\mathcal{O}(\epsilon)$ suppressed compared to that consisting only of $c$-fields, owing to the squeezing. This can be seen explicitly as
\[
\frac{\tilde G^{<ij}_{c\Delta}(t,t';k)}{\tilde G^{<ij}_{cc}(t,t';k)}\sim\begin{cases}
\mathcal{O}\left(\epsilon^3\right) & (i,j)=(1,1) \\
\mathcal{O}(\epsilon) & (i,j)=(1,2), (2,1), (2,2)\,.
\end{cases}
\]
Therefore, our formalism can evaluate IR secular growth which appears in \emph{all} correlation functions in a good approximation even though the operator orderings of calculable correlation functions are uniquely fixed.

\section{Conclusion and discussions} \label{concl}
In this study, we have investigated the dynamics which can describe all IR secular growth in the theory of a minimally coupled massless scalar field on de Sitter background. 
We formulated a systematic way of deriving an effective EoM for IR modes of a massless scalar field with a general potential $V(\phi)$ on de Sitter background. 
We applied our formalism to $\lambda\phi^4$ theory and explicitly derived an effective EoM for IR modes which can describe IR secular growth up to NLO, and showed that NLO IR secular growth could be interpreted as an increase of the statistical variance in a region where UV modes and IR modes are weakly interacting with each other. However, it is yet to be studied how to justify to treat the noise as a classical stochastic noise beyond NLO. We will study this aspect further in our future work.
 In order to apply our formalism to more realistic inflationary models, we need to take into account the backreaction onto the background geometry, which has been neglected in this study. We will also study this aspect in our future work.

\acknowledgments

We would like to thank H. Kitamoto for discussions. T. T. was also supported in part by MEXT Grant-in-Aid for Scientific Research on Innovative Areas, Nos. 17H06357 and 17H06358, and by Grant-in-Aid for Scientific Research  Nos. 26287044 and 15H02087.  

\appendix
\section{The products of mode functions}
In many cases, the momenta of propagators are restricted to $k=\epsilon aH$ due to the bilinear vertex. In this case, the products of mode functions are evaluated as follows.
\begin{align}
\left. \modefn{}{t}{t'}\right|_{k=\epsilon aH}
&=\frac{H^2}{2(\epsilon aH)^3}\left(1+\frac{\epsilon^2}{2}+\frac{i\epsilon^3}{3}+\mathcal{O}\left(\epsilon^4\right)\right)\left(1+i\epsilon \frac{a}{a(t')}\right)e^{-i\epsilon a/a(t')}\,,\label{modefunction1}\\
\left. \dot\Phi_k(t)\Phi^*_k(t')\right|_{k=\epsilon aH}&=\frac{-H^3\epsilon^2}{2(\epsilon aH)^3}e^{i\epsilon}\left(1+i\epsilon \frac{a}{a(t')}\right)e^{-i\epsilon a/a(t')}\,,\label{modefunction2}\\
\mathrm{Re}\left[\left.\modefn{}{t}{t'}\right|_{k=\epsilon aH}\right]&=\frac{H^2}{2(\epsilon aH)^3}f_R(\Delta t)\left(1+\mathcal{O}\left(\epsilon^2\right)\right)\,,\label{modefunction3}\\
\mathrm{Re}\left[\left. \dot\Phi_k(t)\Phi^*_k(t')\right|_{k=\epsilon aH}\right]
&=\mathrm{Re}\left[\left.\modefn{}{t}{t'}\right|_{k=\epsilon aH}\right]\times(-H\epsilon^2)\left(1+\mathcal{O}(\epsilon)\right)\,,\label{modefunction4}\\
\mathrm{Im}\left[\left.\modefn{}{t}{t'}\right|_{k=\epsilon aH}\right]&=\frac{H^2}{2(\epsilon aH)^3}\left(f_I(\Delta t)+\frac{\epsilon^3}{3}f_R(\Delta t)\right)\,,\label{modefunction5}\\
\mathrm{Im}\left[\left. \dot\Phi_k(t)\Phi^*_k(t')\right|_{k=\epsilon aH}\right]&=-\frac{1}{2a^3\epsilon}
\left[f_I(\Delta t)+\epsilon f_R(\Delta t)\right]\,.\label{modefunction6}
\end{align}
Here, $f_R(\Delta t)$ and $f_I(\Delta t)$ are defined by
\begin{align}
\Delta t &\coloneqq t-t' \,,\no\\
f_R(\Delta t)&\coloneqq \remodefn{t}\,,\no\\
f_I(\Delta t)&\coloneqq \immodefn{t}\,, \label{defoff}
\end{align}
which show rapid oscillations for $\Delta t \gg \frac{1}{H}\ln(1/\epsilon)$.
On the other hand, they do not show rapid oscillations for $\Delta t \ll \frac{1}{H}\ln(1/\epsilon)\Leftrightarrow \epsilon a/a(t')\ll 1$, and approximately behave as
\begin{align}
f_R(\Delta t)=1+\mathcal{O}\left(\epsilon^2\right)\,, \quad f_I(\Delta t)=-\frac{\epsilon^3}{3}\left(\frac{a}{a(t')}\right)^3\,.\label{behavioroff}
\end{align}
Therefore, for  $\Delta t \ll \frac{1}{H}\ln(1/\epsilon)$, the products of mode functions can be evaluated approximately as follows:
\begin{align}
\mathrm{Re}\left[\left. \modefn{}{t}{t'}\right|_{k=\epsilon aH}\right]&=\frac{H^2}{2(\epsilon aH)^3}\left(1+\mathcal{O}\left(\epsilon^2\right)\right)\,,\no \\
\mathrm{Re}\left[\left. \dot\Phi_k(t)\Phi^*_k(t')\right|_{k=\epsilon aH}\right]&=-\frac{H^3}{2(\epsilon aH)^3}\times \epsilon^2\left(1+\mathcal{O}\left(\epsilon^2\right)\right)\,,\no\\
\mathrm{Im}\left[\left. \modefn{}{t}{t'}\right|_{k=\epsilon aH}\right]&=\frac{H^2}{2(\epsilon aH)^3}\frac{\epsilon^3}{3}\left(1-\left(\frac{a}{a(t')}\right)^3\right)\left(1+\mathcal{O}\left(\epsilon^2\right)\right)\,,\no\\
\mathrm{Im}\left[\left. \dot\Phi_k(t)\Phi^*_k(t')\right|_{k=\epsilon aH}\right]&=-\frac{1}{2a^3}\left(1+\mathcal{O}\left(\epsilon^2\right)\right)\,.\label{irmode}
\end{align}

\section{Derivation of LO stochastic dynamics} \label{exercise}
In this appendix, by using eq.~\eqref{IRstrength}, we derive LO stochastic EoM described by eq.~\eqref{LOstoc}  as an exercise. First of all, we should mention that the form of eq.~\eqref{unifiedeom} is different from eq.~\eqref{LOstoc}. From eqs.~\eqref{formalsol} and \eqref{Greenfn}, one can see that the following approximation is valid up to LO:  
\begin{equation}
G(T,t)\simeq\frac{-i}{3H}\frac{1}{a^3}\Theta\left(T-t\right)\,.\label{LOret}
\end{equation}
Then, the expression \eqref{formalsol} is the formal solution of the following equation:
\begin{equation}
3H\dirphic=-V'_\mathrm{eff}(\irphic)+\dot\mu+3H\mu+\tilde\xi\,.\label{LOstoc2}
\end{equation}
One may notice that the approximation \eqref{LOret} corresponds to neglecting the second order time derivative term of $\irphic$ from the effective EoM of $\irphic$.

Furthermore, up to LO, $\tilde\xi$ reduces to 
\begin{equation}
\tilde\xi\simeq3H\xi_\phi+\xi_v \,,\label{LOstoc3}
\end{equation}
because there is no oscillatory function in the integrand on the right-hand side of eq.~\eqref{formalsol} up to LO, and as a result $\dot\xi_\phi$ term in $\tilde\xi$ cannot contribute to LO IR secular growth.\footnote{As one can see from eq.~\eqref{LOstoc}, the noise correlation has no oscillatory feature because $\delta(t_1-t_2)$ is replaced by an exponentially decaying function with respect to $|t_1-t_2|$ after coarse-graining.} Combining eq.~\eqref{LOstoc2} and \eqref{LOstoc3}, eq.~\eqref{unifiedeom} reduces to 
\begin{equation}
\dirphic=-\frac{1}{3H}\left(V'_\mathrm{eff}(\irphic)-\dot\mu(\irphic)\right)+\mu(\irphic)+\xi_\phi+\frac{1}{3H}\xi_v\,.\label{LOstoc4}
\end{equation}
Let us consider $\lambda\phi^4$ theory for example.
As we discussed in sec.~\ref{NLOderivation}, only $0$-th order diagrams can contribute to LO IR secular growth. Because there is no $0$-th order $V_\mathrm{eff}$ diagram, LO vertex is only a pure IR vertex $V'(\irphic)=\lambda\irphid\irphic^3/6$. This pure IR vertex corresponds to $l=1,n=4, m=3$ diagram, which satisfies $(n/2)+2l-(m+1)=0$ indeed. Therefore, $V'_\mathrm{eff}(\irphic)$ reduces to $\lambda\irphic^3/6=V'(\irphic)$ and $\mu=0$ in eq.~\eqref{LOstoc} up to LO. $0$-th order noise diagrams are shown in fig.~\ref{snownoise1}. These diagrams correspond to $l=0,n=2,m=0$ diagrams, which also satisfy $(n/2)+2l-(m+1)=0$, and hence contribute to LO IR secular growth. However, by performing straightforward calculation, it turns out that among these diagrams, only the diagram without $\irphid$ fields has $\mathcal{O}\left(\epsilon^0\right)$ values and other diagrams are $\mathcal{O}(\epsilon)$ suppressed. Therefore, $\xi_v$ terms on the right-hand side of eqs.~\eqref{LOstoc3} and \eqref{LOstoc4} are negligible. One can show that the noise diagram in fig.~\ref{snownoise1} without $\irphid$ fields generates the noise correlation $\left<\xi_\phi\xi_\phi\right>$ which is described by eq.~\eqref{LOstoc}.

\section{Consistency check}
As a necessary consistency check, we calculate $\bra0{\phi^<}^2(x)\ket0$ perturbatively with respect to $\lambda$ up to NLO in two ways: firstly in perturbative QFT technique and secondly in our NLO stochastic EoM. Then, we compare these two results with each other. We need to calculate the coefficient of $\lambda(\ln a/a_0)^2$  because we want to calculate first order terms of $\bra0{\phi^<}^2(x)\ket0$ with respect to $\lambda$ up to NLO.

\subsection{QFT calculation}\label{QFTcomp}
\begin{figure}[tbp]
 \centering
  \includegraphics[width=.8\textwidth,trim=80 550 40 80,clip]{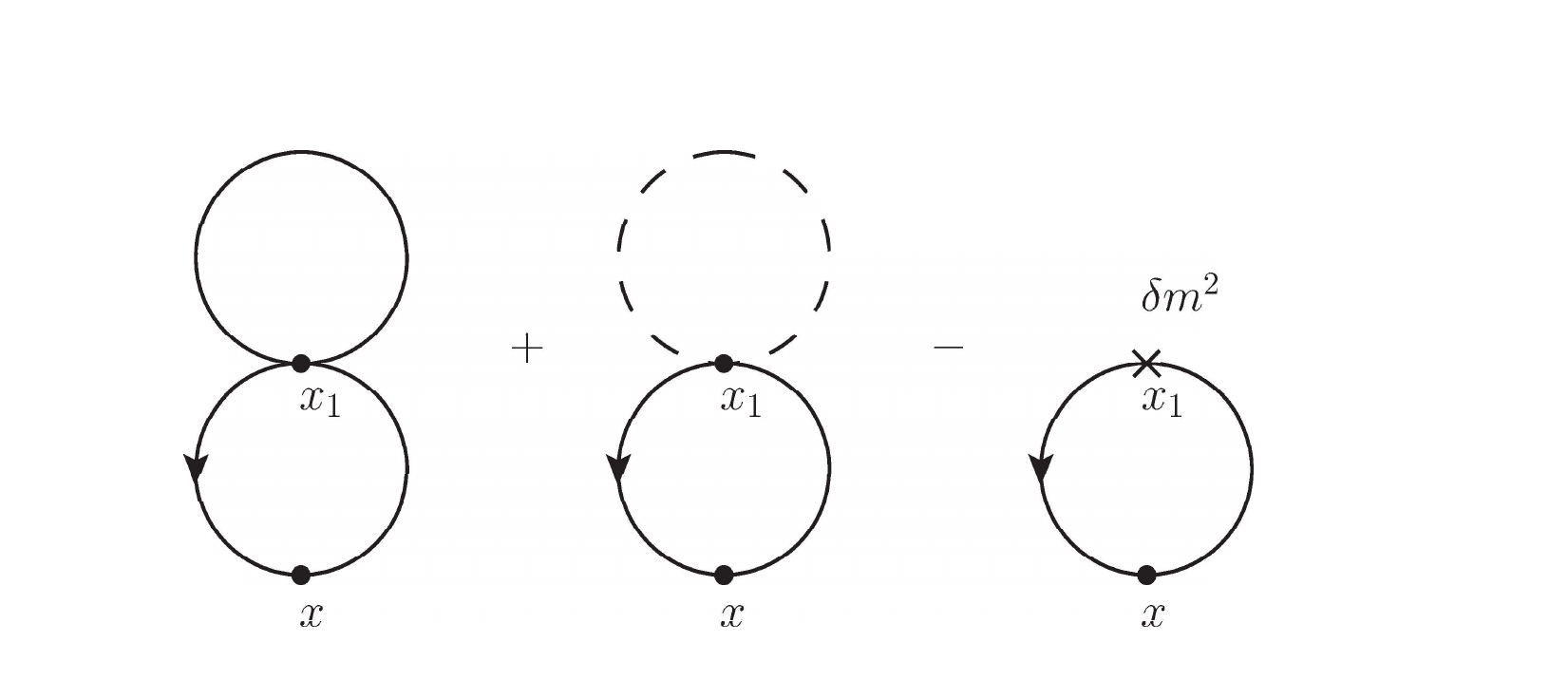}
 \caption{These diagrams are first order corrections with respect to $\lambda$ to $\bra0{\phi^<}^2(x)\ket0$. An effective EoM for IR modes in our formalism can recover IR correlation functions.}
 \label{2-point} 
\end{figure}
First, we perform an usual perturbative QFT calculation up to NLO. 
The first order corrections to $\bra0{\phi^<}^2(x)\ket0$ with respect to $\lambda$ are shown in fig.~\ref{2-point} when the renormalization scheme is chosen so that the contribution from $\delta m^2$ term and that from UV loop integral are canceled out. Then, $\bra0{\phi^<}^2(x)\ket0$ is evaluated up to first order with respect to $\lambda$ as 
\begin{align}
\bra0{\phi^<}^2(x)\ket0
&\simeq\int^{\epsilon aH}_{\epsilon a_0H}\frac{\mathrm{d}^3 k}{(2\pi)^3}\mathrm{Re}[\Phi_k(t)\Phi^*_k(t)]\no\\
&\quad-i\lambda\int^{t}_{t_0}\mathrm{d}t'\,{a^3(t')}\frac{H^2}{4\pi^2}\ln{\frac{a(t')}{a_0}}\int^{\epsilon aH}_{\epsilon a_0H}\frac{\mathrm{d}^3 k}{(2\pi)^3}2i\,\mathrm{Im}[\Phi_k(t)\Phi^*_k(t')]\mathrm{Re}[\Phi_k(t)\Phi^*_k(t')]\,,\label{qft2}
\end{align}
where we neglect $\mathcal{O}(\epsilon)$ terms.
Next, by using the following equations
\begin{subequations}
\label{QFT1}
\begin{align}
\mathrm{Re}[\Phi_k(t)\Phi_k^*(t')]_{k\leq\epsilon aH}&=\frac{H^2}{2k^3}\left(\frac{k}{a(t')H}\sin\biggl(\frac{k}{a(t')H}\biggl)+\cos \biggl(\frac{k}{a(t')H}\biggl)\right)\Biggl(1+\mathcal{O}\biggl(\Bigl(\frac{k}{aH}\Bigl)^2\biggl)\Biggl)\,,\\
\mathrm{Im}[\Phi_k(t)\Phi_k^*(t')]_{k\leq\epsilon aH}&=\frac{H^2}{2k^3}\Biggl[\Biggl(\frac{k}{a(t')H}\cos \biggl(\frac{k}{a(t')H}\biggl)-\sin\biggl(\frac{k}{a(t')H}\biggl)\Biggl)\no\\
&\quad+\frac{1}{3}\biggl(\frac{k}{aH}\biggl)^3\Biggl(\frac{k}{a(t')H}\sin\biggl(\frac{k}{a(t')H}\biggl)+\cos \biggl(\frac{k}{a(t')H}\biggl)\Biggl)\Biggl]\Biggl(1+\mathcal{O}\biggl(\Bigl(\frac{k}{aH}\Bigl)^2\biggl)\Biggl)\,,
\end{align}
\end{subequations}
and neglecting higher order terms with respect to $(k/aH)$, the first order terms of $\bra0{\phi^<}^2(x)\ket0$ with respect to $\lambda$ are given by
\begin{align}
&\frac{\lambda H^3}{16\pi^4}\left\{\int^{t}_{t_0}\mathrm{d}t'
\ln\frac{a(t')}{a_0}\left[F\left(\frac{\epsilon a_0}{a(t')}\right)
-F\left(\frac{\epsilon a}{a(t')}\right)\right]+\frac{1}{3}\int^{t}_{t_0}\mathrm{d}t'\left(\frac{a(t')}{a}\right)^3\ln\frac{a(t')}{a_0}\int^{\frac{\epsilon a}{a(t')}}_{\frac{\epsilon a_0}{a(t')}}\frac{\mathrm{d}K}{K}(K\sin K+\cos K)^2\right\}\,.
\label{2pointNLO}
\end{align}
Here, we define new dimensionless variable $K$ by $K\coloneqq k/a(t')H$ and function $F$ by
\begin{align}
F(x)\coloneqq\int^{\infty}_x\frac{\mathrm{d}K}{K^4}(K\sin K+\cos K)(K\cos K-\sin K)\,.\label{defF}
\end{align}
From the definition of $K$, $K\geq1$ corresponds to UV modes at time $t'$, and $K\leq1$ corresponds to IR modes at time $t'$. Let us calculate the first term in eq.~\eqref{2pointNLO}.  
$F(\epsilon a/a(t'))$ decays with oscillating for $t_0+\ln(1/\epsilon)/H\lesssim t'\leq t$, which is the dominant part of the time integral region, and hence does not contribute to $\lambda(\ln a/a_0)^2$ terms in $\bra0{\phi^<}^2(x)\ket0$. On the other hand, $F(\epsilon a_0/a(t'))$ can contribute to $\lambda(\ln a/a_0)^2$ terms in $\bra0{\phi^<}^2(x)\ket0$. $F(\epsilon a_0/a(t'))$ can be expanded analytically as 
\begin{align}
F\left(\frac{\epsilon a_0}{a(t')}\right)=\frac{1}{3}\left[\ln\left(\frac{\epsilon a_0}{a(t')}\right)+\ln 2+\gamma-\frac{7}{3}+\mathcal{O}\left(\left(\frac{\epsilon a_0}{a(t')}\right)^2\right)\right]\,.
\end{align}
$\mathcal{O}\left(\epsilon a_0/a(t')\right)$ terms can be neglected because an inequality $\frac{\epsilon a_0}{a(t')}<\epsilon\ll1$ is always satisfied in the integral region $t_0\leq t'\leq t$. Here, $\gamma$ is an Euler's constant. Therefore,
\begin{align}
&\int^{t}_{t_0}\mathrm{d}t'\ln\frac{a(t')}{a_0}\left[F\left(\frac{\epsilon a_0}{a(t')}\right)-F\left(\frac{\epsilon a}{a(t')}\right)\right]\no\\
&=-\frac{1}{9H}\left(\ln\frac{a}{a_0}\right)^3+\frac{1}{6H}\left(-\ln\frac{1}{\epsilon}+\ln2+\gamma-\frac{7}{3}\right)(\ln \frac{a}{a_0})^2+\mathcal{O}\left(\ln\frac{a}{a_0}\right)\,.\label{2pointNLOresult1}
\end{align}
Next, let us calculate the second term in eq.~\eqref{2pointNLO}. This term is calculated as
\begin{align}
&\int^{t}_{t_0}\mathrm{d}t'\biggl(\frac{a(t')}{a}\biggl)^3\ln\frac{a(t')}{a_0}\int^{\frac{\epsilon a}{a(t')}}_{\frac{\epsilon a_0}{a(t')}}\frac{\mathrm{d}K}{K}(K\sin K+\cos K)^2\no\\
&=\int^{t}_{t_0}\mathrm{d}t'\biggl(\frac{a(t')}{a}\biggl)^3\ln\frac{a(t')}{a_0}\cdot\Biggl\{\frac{1}{2}\ln\frac{a}{a_0}+\frac{1}{2}\left[\mathrm{Ci}\left(\frac{2\epsilon a}{a(t')}\right)-\mathrm{Ci}\left(\frac{2\epsilon a_0}{a(t')}\right)\right]+\frac{5}{8}\left[\cos \left(\frac{2\epsilon a}{a(t')}\right)-\cos \left(\frac{2\epsilon a_0}{a(t')}\right)\right]\no\\
& \qquad +\frac{1}{4}\left[\left(\frac{2\epsilon a}{a(t')}\right)^2-\left(\frac{2\epsilon a_0}{a(t')}\right)^2\right]-\frac{1}{4}\left[\frac{2\epsilon a}{a(t')}\sin\left(\frac{2\epsilon a}{a(t')}\right)-\frac{2\epsilon a_0}{a(t')}\sin\left(\frac{2\epsilon a_0}{a(t')}\right)\right]\Biggl\}\,. \label{NLOsecondterm}
\end{align}
Here, $\mathrm{Ci}(x)$ is defined by 
\begin{equation*}
\mathrm{Ci}(x)\coloneqq\int^x_\infty\mathrm{d}x\frac{\cos(x)}{x}\,.
\end{equation*}
Because the powers of $a(t')$ in the first factor of the integrand in eq.~\eqref{NLOsecondterm} are positive, $(\ln a/a_0)^2$ terms appear from this integral only when this positive powers of $a(t')$ are canceled or the powers of $\ln(a/a_0)$ in the integrand becomes equal to or more than two.  For this reason, the terms in the bracket $\{\cdots\}$ in eq.~\eqref{NLOsecondterm} which contribute to $(\ln a/a_0)^2$ terms are $\ln(a/a_0)$ term and $\mathrm{Ci}(\frac{2\epsilon a_0}{a(t')})$ term. Indeed, $\mathrm{Ci}(x)$ can be expanded for small argument $x$ as
\begin{equation*}
\mathrm{Ci}(x)=\ln(x)+\gamma+\mathcal{O}\left(x^2\right)\,,
\end{equation*}
which shows logarithmic divergence for small $x$ and this divergence becomes the origin of $(\ln a/a_0)^2$ terms in eq.~\eqref{NLOsecondterm}.\footnote{Although $\mathrm{Ci}(x)$ is logarithmically divergent as $\ln x$, it decays rapidly for $x\geq1$. This is why $\mathrm{Ci}(\frac{2\epsilon a}{a(t')})$ does't contribute to $(\ln a/a_0)^2$ terms in eq.~\eqref{NLOsecondterm}.}
Using this equation, eq.~\eqref{NLOsecondterm} is calculated as 
\begin{align}
&\int^{t}_{t_0}\mathrm{d}t'\biggl(\frac{a(t')}{a}\biggl)^3\ln\frac{a(t')}{a_0}\int^{\frac{\epsilon a}{a(t')}}_{\frac{\epsilon a_0}{a(t')}}\frac{\mathrm{d}K}{K}(K\sin K+\cos K)^2
=\frac{1}{3H}\left(\ln\frac{a}{a_0}\right)^2+\mathcal{O}\left(\ln\frac{a}{a_0}\right)\,. \label{2pointNLOresult2}
\end{align}
Therefore, by substituting eqs.\eqref{2pointNLOresult1} and \eqref{2pointNLOresult2} into eq.~\eqref{2pointNLO}, and using eq.~\eqref{qft2}, one obtains
\begin{align}
&\bra0{\phi^<}^2(x)\ket0\no\\
&=\frac{H^2}{4\pi^2}\ln\frac{a}{a_0}+\lambda H^2\Biggl[-\frac{1}{144\pi^4}
\left(\ln\frac{a}{a_0}\right)^3+\frac{1}{96\pi^4}\left(-\ln\frac{1}{\epsilon}+\ln2+\gamma-\frac{5}{3}\right)\left(\ln\frac{a}{a_0}\right)^2+\mathcal{O}\left(\ln\frac{a}{a_0}\right)\Biggl]+\mathcal{O}\left(\lambda^2\right)\,. \label{QFTresult}
\end{align}
\subsection{Stochastic calculation}
Next, we calculate $\bra0{\phi^<}^2(x)\ket0$ perturbatively with respect to $\lambda$ up to NLO, by using our NLO stochastic eq.~\eqref{nlosto2}. It is convenient to rewrite eq.~\eqref{nlosto2} in the form of Fokker-Planck equation in evaluating correlation functions. When we derive the corresponding Fokker-Planck equation, we need to define the integral of the stochastic noise because the amplitude of the stochastic noise depends on field $\irphic$. Since the physical origin of this field dependence of the stochastic noise is that UV modes feel the background value of IR modes $\irphic$, we should adopt the It\^o integral from causality.
Therefore, we treat eq.~\eqref{nlosto2} as the It\^o type Langevin equation, and the corresponding Fokker-Planck equation can be obtained as\cite{Risken1984}
\begin{align}
\dot\rho(\irphic,\irvc,t)&=\frac{1}{2}\frac{\der^2}{\der\irphic^2}\left(N\rho\right)-\frac{\der}{\der\irphic}\left(\irvc\rho\right)+\frac{\der}{\der\irvc}\left(3H\irvc+\frac{\lambda}{6}\irphic^3\rho\right)\,. \label{NLOFP}
\end{align}
By multiplying both sides of eq.~\eqref{NLOFP} by $\irphic^n\irvc^m$ and performing partial integral with respect to $\irphic$ and $\irvc$, one can get the following recurrence relation:
\begin{align}
\frac{\der}{\der t}\left<\irphic^n\irvc^m\right>
&=\left[\frac{\lambda Hn(n-1)}{8\pi^2}\left(l(\epsilon)-\frac{1}{9}\right)-3mH\right]\left<\irphic^n\irvc^m\right>\no\\
&\quad+\frac{n(n-1)H^3}{8\pi^2}\left<\irphic^{n-2}\irvc^m\right>+n\left<\irphic^{n-1}\irvc^{m+1}\right>-\frac{m\lambda}{6}\left<\irphic^{n+3}\irvc^{m-1}\right>\,.\label{diffeq}
\end{align}
Here, initial conditions are given by
\begin{align}
\left<\irphic^n\irvc^m\right>=0 \qquad \mathrm{for} \quad t=t_0\,, \label{initial}
\end{align}
and $n$ and $m$ are non-negative integers.
Next, we calculate $\left<\irphic^2(x)\right>$ up to first order with respect to $\lambda$ by using eqs.~\eqref{diffeq} and \eqref{initial}. Substituting $(n,m)=(2,0) , (1,1) , (0,2)$ into eq.~\eqref{diffeq}, one obtains
\begin{align}
\frac{\der}{\der t}\left<\irphic^2\right>&=\frac{H^3}{4\pi^2}+\frac{\lambda H}{4\pi^2}\left(l(\epsilon)-\frac{1}{9}\right)\left<\irphic^2\right>+2\left<\irphic \irvc\right>\,,\label{(2,0)diffeq}\\
\frac{\der}{\der t}\left<\irphic \irvc\right>&=-3H\left<\irphic\irvc\right>+\left<\irvc^2\right>-\frac{\lambda}{6}\left<\irphic^4\right>\,,\label{(1,1)diffeq}\\
\frac{\der}{\der t}\left<\irvc^2\right>&=-6H\left<\irvc^2\right>-\frac{\lambda}{3}\left<\irphic^3\irvc\right>\,. \label{(0,2)diffeq}
\end{align}
From eq.~\eqref{(0,2)diffeq}, the 0-th order term of $\left<\irvc^2\right>$ with respect to $\lambda$ satisfies
\begin{equation*}
\frac{\der}{\der t}\left<\irvc^2\right>=-6H\left<\irvc^2\right>\,.
\end{equation*}
Since the solution of this differential equation with an intial condition \eqref{initial} is only a trivial solution $\left<\irvc^2\right>=0$, $\left<\irvc^2\right>$ is, at least, first order in $\lambda$. By using this fact and eq.~\eqref{(1,1)diffeq}, it is also obvious that $\left<\irphic\irvc\right>$ is, at least, first order in $\lambda$. Using the fact $\left<\irphic\irvc\right>\sim\mathcal{O}(\lambda)$ and eq.~\eqref{(0,2)diffeq}, it turns out that $\left<\irvc^2\right>$ is, at least, second order in $\lambda$.
Using these facts and $\left<\irphic^4\right>=3\left<\irphic^2\right>^2+\mathcal{O}(\lambda)$, eqs.~\eqref{(2,0)diffeq} and \eqref{(1,1)diffeq} reduce to
\begin{align}
\frac{\der}{\der t}\left<\irphic^2\right>&=\frac{H^3}{4\pi^2}+\frac{\lambda H}{4\pi^2}\left(l(\epsilon)-\frac{1}{9}\right)\left<\irphic^2\right>+2\left<\irphic \irvc\right>\,,\label{phiphidiffeq}\\
\frac{\der}{\der t}\left<\irphic\irvc\right>&=-3H\left<\irphic\irvc\right>-\frac{\lambda}{2}\left<\irphic^2\right>^2+\mathcal{O}\left(\lambda^2\right)\,.\label{phivdiffeq}
\end{align}
Let us put the ansatz as follows:
\begin{align}
\left<\irphic^2\right>&=\frac{H^2}{4\pi^2}\ln\frac{a}{a_0}+\lambda H^2\left(c_1\left(\ln\frac{a}{a_0}\right)^3+c_2\left(\ln\frac{a}{a_0}\right)^2+\mathcal{O}\left(\ln\frac{a}{a_0}\right)\right)+\mathcal{O}\left(\lambda^2\right)\,,\label{phiphiansatz}\\
\left<\irphic\irvc\right>&=\lambda H^3\left(d_1\left(\ln\frac{a}{a_0}\right)^2+d_2\ln\frac{a}{a_0}+d_3+\mathcal{O}\left(\frac{a_0}{a}\right)\right)+\mathcal{O}\left(\lambda^2\right)\,,\label{phivansatz}
\end{align}
where $c_1,\cdots,\, d_3$ are constants. Substituting eqs.~\eqref{phiphiansatz} and \eqref{phivansatz} into eqs.~\eqref{phiphidiffeq} and \eqref{phivdiffeq}, and focusing on the coefficients of the first order terms with respect to $\lambda$, one obtains
\begin{align*}
3c_1\left(\ln\frac{a}{a_0}\right)^2+2c_2\ln\frac{a}{a_0}+\mathcal{O}\left(\left(\ln\frac{a}{a_0}\right)^0\right)
&=2d_1\left(\ln\frac{a}{a_0}\right)^2+\left[\frac{l(\epsilon)-1/9}{16\pi^4}+2d_2\right]\ln\frac{a}{a_0}+\mathcal{O}\left(\left(\ln\frac{a}{a_0}\right)^0\right)\,,\\
2d_1\ln\frac{a}{a_0}+d_2
&=-\left(3d_1+\frac{1}{32\pi^4}\right)\left(\ln\frac{a}{a_0}\right)^2-3d_2\ln\frac{a}{a_0}+\mathcal{O}\left(\left(\ln\frac{a}{a_0}\right)^0\right)\,.
\end{align*}
From the above two identities, the following equations follow:
\begin{align*}
3c_1=2d_1\,, \quad 2c_2=2d_2+\frac{l(\epsilon)-1/9}{16\pi^4}\,,\quad d_1=-\frac{1}{96\pi^4}\,, \quad 3d_2=-2d_1\,.
\end{align*}
Therefore, one gets
\begin{align*}
d_1&=-\frac{1}{96\pi^4}\,, \quad d_2=\frac{1}{144\pi^4}\,, \quad c_1=-\frac{1}{144\pi^4}\,, \\
c_2&=\frac{1}{32\pi^4}\left(l(\epsilon)+\frac{1}{9}\right)=\frac{1}{96\pi^4}\left(-\ln\frac{1}{\epsilon}+\ln2+\gamma-\frac{5}{3}\right)\,,
\end{align*}
which results in
\begin{align}
\left<\irphic^2\right>
&=\frac{H^2}{4\pi^2}\ln\frac{a}{a_0}+\lambda H^2\left[
-\frac{1}{144\pi^4}
 \left(\ln\frac{a}{a_0}\right)^3+\frac{1}{96\pi^4}
    \left(-\ln\frac{1}{\epsilon}+\ln2+\gamma-\frac{5}{3}\right)
    \left(\ln\frac{a}{a_0}\right)^2+\mathcal{O}\left(\ln\frac{a}{a_0}\right)\right]\,.
\end{align}
This result coincides with the result eq.~\eqref{QFTresult} which is calculated by perturbative QFT technique.

\section{Collinear singularity} \label{colliapp}
\begin{figure}[tbp]
 \centering
  \includegraphics[width=.7\textwidth,trim=70 500 70 130,clip]{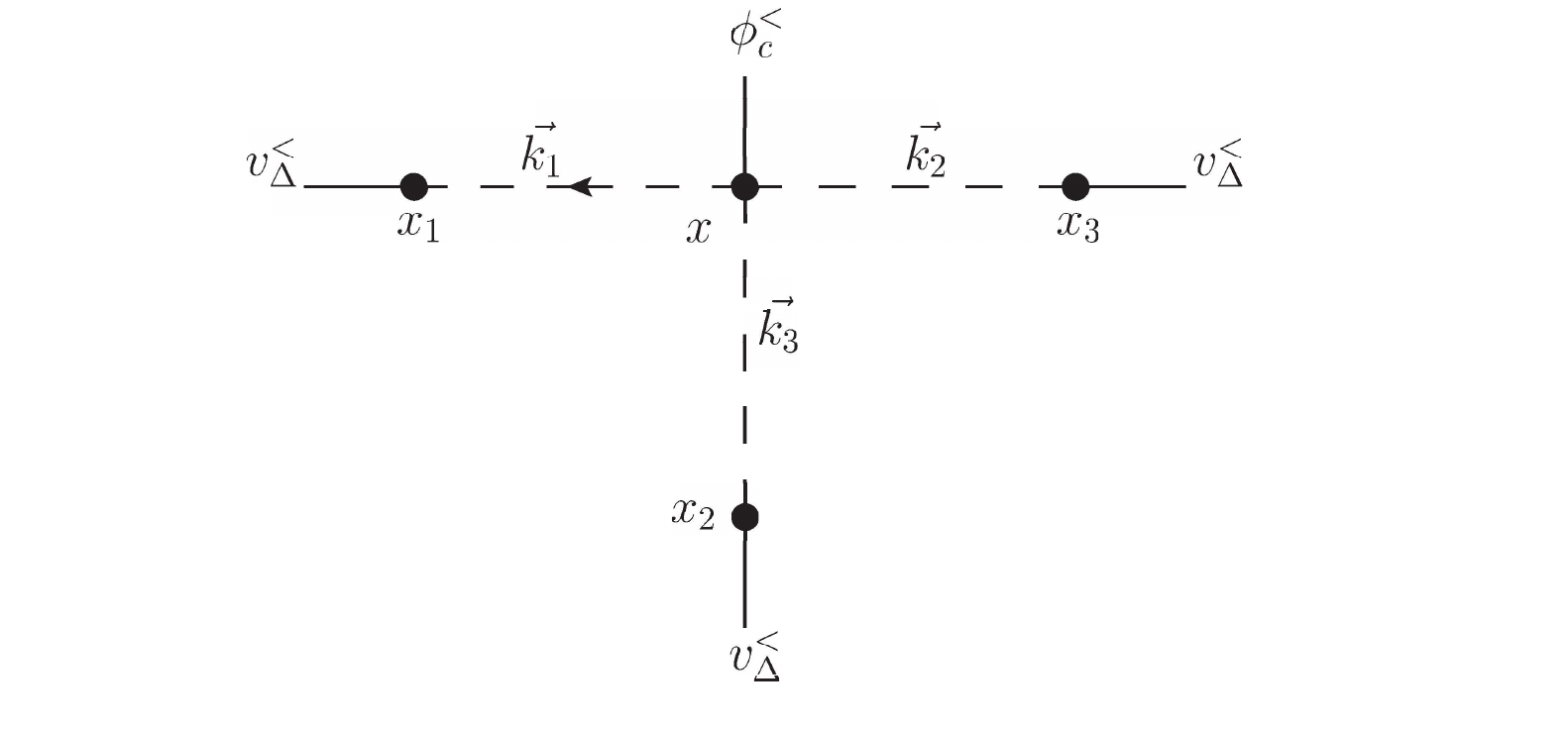}
 \caption{This diagram is diagram (3)-(a) in fig.~\ref{NLOcan2}. In order to show approximate locality of diagrams shown in fig.~\ref{NLOcan2}, we need to prove the approximate locality of this diagram for example.}
 \label{collinear}
\end{figure}

At the beginning of the sec.~\ref{estimation}, we assumed that $i\Gamma$ is approximately local from a coarse-graining point of view. In this appendix, by evaluating diagram shown in fig.~\ref{collinear}, we show that this assumption is true at least for first order diagrams in renormalizable scalar field theories without derivative interactions. Here, first order means the order with respect to the coupling constant. Since this diagram corresponds to diagram (3)-(a) in fig.~\ref{NLOcan2}, we denote the contribution of this diagram to $i\Gamma$ as $i\Gamma_\mathrm{(3)-(a)}$. 
For simplicity, we replace $\irphic\left(\vec x,t\right)$ by $\irphic\left(\vec{x}_1,t\right)$.\footnote{More rigorously, we need to evaluate $i\Gamma_\mathrm{(3)-(a)}$ without this replacement. 
If we do so, the calculation becomes more complicated, but still we can show the approximate locality. Note that the calculation performed here directly applies to the discussion of collinear singularity in $\phi^3$ theory.} As was discussed in sec.~\ref{NLOderivation}, this replacement is valid regarding the most dominant part of $\irphic$ at a late time. Then, $i\Gamma_\mathrm{(3)-(a)}$ is evaluated as
\begin{align}
&i\Gamma_\mathrm{(3)-(a)}=\volint{x_1}\,a^3(t_1)\irvd(x_1)\volint{x_2}\,a^3(t_2)\irvd(x_2)\volint{x_3}\,a^3(t_3)\irvd(x_3)N_\mathrm{(3)-(a)}\,,\\
N_\mathrm{(3)-(a)}
&\sim\lambda H\int^{t_m}\mathrm{d}t\,a^3\irphic(\vec x_1,t)\modeint{2}\modeint{3}e^{i\vec k_2\cdot(\vec x_2-\vec x_1)}e^{i\vec k_3\cdot(\vec x_3-\vec x_1)}\no\\
&\quad \times\delta\left(t_1-t_c\left(\bigl|\vec k_2+\vec k_3\bigr|\right)\right)\delta(k_2-\ah{2})\delta(k_3-\ah{3})\frac{1}{(\epsilon a(t_1))^3}\frac{1}{(\epsilon a(t_2))^2}\frac{1}{(\epsilon a(t_3))^2}\no\\
&\quad \times\left\{\left(f_I(\Delta t_1)+\frac{\epsilon^3}{3}f_R(\Delta t_1)\right)f_R(\Delta t_2)f_R(\Delta t_3)+(t_1\leftrightarrow t_2)+(t_1\leftrightarrow t_3)\right\}\label{3-a}\,,
\end{align}
where we define $t_m\coloneqq \mathrm{min}(t_1,t_2,t_3)$ and $\Delta t_i\coloneqq t_i-t$ for $i=1,2,3$, and $f_\mathrm{I}$ and $f_\mathrm{R}$ were introduced in eq.~\eqref{defoff}. First, we show that the integrand of eq.~\eqref{3-a} exponentially decays for $t_m-t\gtrsim\ln(1/\epsilon)/H$. For $t_m-t\gtrsim\ln(1/\epsilon)/H$, $f_R(\Delta t_i)$ and $f_I(\Delta t_i)$ behave as
\begin{align}
f_R(\Delta t_i)\simeq\epsilon e^{H\Delta t_i}\sin(\epsilon e^{H\Delta t_i})\,, \quad f_I(\Delta t_i)\simeq\epsilon e^{H\Delta t_i}\cos(\epsilon e^{H\Delta t_i})\,.\label{oscf}
\end{align}
Substituting this eq.~\eqref{oscf} into eq.~\eqref{3-a} and after some tedious computation, the $t$-dependent part in eq.~\eqref{3-a} becomes
\begin{align}
\frac{1}{H}\int^{\epsilon\frac{a_m}{a_0}}_{\epsilon}\mathrm{d}y\,\frac{1}{y}\irphic(\vec x_1,y)
&\Biggl\{\cos\biggl((b_1+b_2-b_3)y\biggl)+\cos\biggl((b_1-b_2+b_3)y\biggl)\no\\
&\quad+\cos\biggl((b_1-b_2-b_3)y\biggl) -3\cos\biggl((b_1+b_2+b_3)y\biggl)\Biggl\}\,,\label{3-a-1}
\end{align}
where $y$, $a_m$, and $b_i \ (i=1,2,3)$ are defined by
\begin{align*}
y&\coloneqq\epsilon \frac{a_m}{a}\,, \quad a_m\coloneqq a(t_m)\,, \quad
b_i\coloneqq\frac{a(t_i)}{a_m}\,, \quad \mathrm{for}\quad i=1,2,3\,.
\end{align*}
Note that $y\geq1$ is equivalent to $t_m-t\geq\ln(1/\epsilon)/H$. Because the time dependence of $\irphic(\vec x_1,t)$ is typically the powers of $\ln a$, eq.~\eqref{3-a-1} converges for $y\geq1$ unless the arguments of cosine function are zero, that is, $b_1\pm b_2\pm b_3=0$.\footnote{One can see this fact more explicitly by performing partial integral.} 
These conditions can be satisfied only when $\vec k_1\,\para\!\!
\vec k_2\,\para\!\!
\vec k_3$ is satisfied, and hence correspond to collinear singularities. Intuitively it is obvious that these singularities do not contribute to the integral in eq.~\eqref{3-a-1} because as the magnitude of $y$ gets larger and larger, the region of the phase space of momenta gets narrower and narrower in which the integrand does not oscillate. In order to see this behavior more precisely, let us consider the case $t_1\geq t_2\geq t_3=t_m$ and derive the condition for almost no oscillation of $\cos\bigl((b_1-b_2-b_3)y\bigr)$. Note that one can assume $t_1\geq t_2\geq t_3=t_m$
without loss of generality because $i\Gamma_\mathrm{(3)-(a)}$ is completely symmetric with respect to $x_1$, $x_2$, and $x_3$. By using the angle $\theta$ between $\vec k_2$ and $\vec k_3$, $b_1-b_2-b_3$ is expanded with respect to $\theta$ as 
\begin{align}
b_1-b_2-b_3&=\frac{\bigl|\vec k_2+\vec k_3\bigr|-k_2-k_3}{\ah{3}}=\frac{-a(t_1)a(t_2)}{2(a(t_2)+a(t_3))^2}\cdot\theta^2\left(1+\mathcal{O}\left(\theta^2\right)\right)\sim\mathcal{O}(1)\times\theta^2\,.\label{thetaexp}
\end{align}
Note that $b_1-b_2-b_3=0\Leftrightarrow\theta=0$. It should be also noted that $a(t_1)a(t_2)/(a(t_2)+a(t_3))^2$ is $\mathcal{O}(1)$ quantity as a consequence of momentum conservation, and hence we omit this factor from now on. From the above equation, one can see that $\cos\bigl((b_1-b_2-b_3)y\bigl)$ does not oscillate for $y\geq1$ only when
\begin{equation}
\theta^2y\lesssim1\Leftrightarrow |\,\theta\,|\lesssim\frac{1}{\sqrt y}\leq1\,.
\end{equation}
Note that the assumption $|\,\theta\,|\lesssim1$ is always satisfied in the situation under consideration: $y\geq1$. Next, we integrate over $\theta$ for $|\,\theta\,|\leq1/\sqrt y\leq1$.
For small $\theta$, the $\theta$-dependent terms in the integrand in eq\eqref{3-a} are $\sin\theta$ and $\cos(\theta^2y)$. The first one, $\sin\theta$, comes from the integral measure. The second one, $\cos(\theta^2y)$, comes from $\cos\bigl((b_1-b_2-b_3)y\bigl)$. Then, the integration over $\theta$ for $|\,\theta\,|\lesssim1/\!\!\sqrt y\leq1$ becomes
\begin{equation}
\int^{1/\!\!\sqrt y}_0\mathrm{d}\theta \,\sin\theta\cos(\theta^2y)
\simeq\int^{1/\!\!\sqrt y}_0\mathrm{d}\theta \, \theta\sim\frac{1}{y}\,,
\end{equation}
which results in $1/y$ suppression. Therefore, the collinear singularities cannot contribute to the integral, and the integrand of eq.~\eqref{3-a} exponentially decays for $t_m-t\gtrsim\ln(1/\epsilon)/H$.

Next, we show that $N_\mathrm{(3)-(a)}$ exponentially decays for $t_i-t_m\gtrsim1/H,\ i=1,2,3$. If we assume that $t_1\sim t_2\gg t_3(\Leftrightarrow k_1\sim k_2\gg k_3)$, by using $t_c(|\vec k_2+\vec k_3|)\simeq t_c(k_2)$ and $a(t_1)/a\geq a(t_1)/a(t_3)\gg1$, $N_\mathrm{(3)-(a)}$ can be evaluated as
\begin{align}
N_\mathrm{(3)-(a)}&\sim\frac{\lambda H^4a(t_3)}{a(t_1)}\joah{2}{x_1}{x_2}\joah{3}{x_1}{x_3}\delta(t_1-t_2)\no\\
&\quad\times\int^{\epsilon\frac{a(t_3)}{a_0}}_{\epsilon}\mathrm{d}y\,\irphic(\vec x_1,y)\frac{1}{y}\left[\cos\left(\frac{a(t_1)}{a(t_3)}y\right)+\frac{\epsilon^3}{3}\sin\left(\frac{a(t_1)}{a(t_3)}y\right)\right]\sin\biggl(\frac{a(t_1)}{a(t_3)}y\biggl)\biggl(\sin y+\frac{\cos y}{y}\biggl)\,.
\end{align}
For $(t_1-t_3)\sim (t_2-t_3)\gtrsim1/H$, the coefficient of the integral decays while the integrand oscillates rapidly. Therefore, $N_\mathrm{(3)-(a)}$ exponentially decays for $t_i-t_m\gtrsim1/H,\ i=1,2,3$.

Spatial approximate locality is also ensured because of the factors $j_0(\ah{i}|\vec x_i-\vec{x}|)$ for $i=1,2,3$ in the integrand, which lead to exponential decay for $|\vec x_i-\vec{x}|\gg1/(\epsilon aH)$ for $i=1,2,3$ after time coarse-graining. 
 
From the above discussions, we can conclude that $i\Gamma_\mathrm{(3)-(a)}$ are approximately local. By similar discussions, we can also show that other diagrams shown in fig.~\ref{NLOcan2} are also approximately local.

\bibliography{(JCAP2017-1)submitv4.bib}

\end{document}